\documentclass[aps,prb,twocolumn,superscriptaddress]{revtex4}
\usepackage{amsmath}
\usepackage{graphicx}
\usepackage{amsfonts}
\usepackage{bbold}
\usepackage{color}
\usepackage{epstopdf}

\begin{document}

\title{Chiral Y-junction of Luttinger liquid wires at strong coupling:      
fermionic representation}
\author{D.N. Aristov}
\affiliation{``PNPI'' NRC ``Kurchatov Institute'', Gatchina 188300, Russia}
\affiliation{Institute for Nanotechnology, Karlsruhe Institute of Technology, 76021
Karlsruhe, Germany }
\affiliation{Department of Physics, St.Petersburg State University, Ulianovskaya 1,
St.Petersburg 198504, Russia}
\author{P. W\"olfle}
\affiliation{Institute for Nanotechnology, Karlsruhe Institute of Technology, 76021
Karlsruhe, Germany }
\affiliation{Institute for Condensed Matter Theory, Karlsruhe Institute of Technology,
76128 Karlsruhe, Germany}
\date{\today}

\begin{abstract}
We calculate the conductances of a three-terminal junction set-up of
spinless Luttinger liquid wires threaded by a magnetic flux, allowing for
different interaction strength $g_{3}\neq g$ in the third wire. We employ
the fermionic representation in the scattering state picture, allowing for a
direct calculation of the linear response conductances, without the need of
introducing contact resistances at the connection points to the outer ideal
leads. The matrix of conductances is parametrized by three variables. For
these we derive coupled renormalization group (RG) equations, by summing up
infinite classes of contributions in perturbation theory. The resulting
general structure of the RG equations may be employed to describe junctions with
an arbitrary number of wires and arbitrary interaction strength in each
wire. The fixed point structure of these equations (for the chiral
Y-junction) is analyzed in detail. For repulsive interaction ($g,g_{3}>0$)
there is only one stable fixed point, corresponding to the complete
separation of the wires. For attractive interaction ($g<0$ and/or $g_{3}<0$)
four fixed points are found, the stability of which depends on the interaction
strength. We confirm our previous weak-coupling result of lines of fixed
points for special values of the interaction parameters reaching into the
strong coupling domain. We find new fixed points not discussed before, even
at the symmetric line $g=g_{3}$, at variance with the results of Oshikawa et
al.  The pair tunneling phenomenon conjectured by the latter authors is not
found by us.
\end{abstract}

\maketitle

\section{Introduction}

Electric circuits of one-dimensional quantum wires are described by laws
quite different from those of corresponding classical systems. These laws
are non-local, often quantized, and are generally difficult to derive. An
important first step is the exploration of the properties of junctions of
wires. The linear response conductance of a (symmetric) two-terminal
junction in the limit of zero temperature has been found to be either zero
(repulsive) or unity (attractive interaction) in units of the conductance
quantum \cite{Kane1992,Furusaki1993}. Most of the existing theories employ
the exactly solvable Tomonaga-Luttinger liquid (TLL) model, in which
backward scattering is neglected \cite{GiamarchiBook}. Moreover, for
simplicity the fermions are assumed to be spinless, in which case backward
scattering only gives rise to a renormalization of the forward scattering
amplitude. Furthermore, most of the early works on the problem of two
TLL-wires connected by a junction employed the method of bosonization, or,
for special values of the interaction parameter, used a mapping on to
exactly solvable models\cite{Weiss1995,Fendley1995}. 

An alternative formulation using a purely fermionic representation in terms
of scattering states and a renormalization group (RG) approach for the
conductance has been pioneered by \cite{Yue1994}. The latter formulation has
several advantages over the bosonic approach. Perhaps the most important
difference is that it allows to describe the physically relevant system of a
quantum wire (with interaction between the fermions) adiabatically coupled
to reservoirs (with negligible interaction), whereas the bosonic approach is
formulated for the infinite system. The effect of a finite wire length has only been convincingly derived for a clean wire \cite{Maslov1995,Ponomarenko1996,Safi1995}
; in the case of wires connected by a junction a plausible, but unproven ad hoc procedure has been proposed \cite{Oshikawa2006}. As will be discussed below we have reasons to suspect that the latter procedure is not always correct. The fermionic scattering state approach is generally applicable,
allowing for the study of wires with two impurities \cite{Polyakov2003}, of
multi-lead junctions \cite{Das2004,Lal2002} as well as the inclusion of spin, 
of backward scattering, \cite{Yue1994} of out of equilibrium situations, \cite{Devillard2008} point contact in a quantum spin Hall insulator \cite{Teo2009}. The original
model calculation \cite{Yue1994} has been limited to weak coupling. It may
be shown, however, that the method can be extended to work in the strong
coupling regime as well, by summing up an infinite class of leading
contributions in perturbation theory \cite{Aristov2009}. In all cases
considered so far it has been found that the method provides results in
agreement with any available exact results, which builds confidence in its
validity. As shown in \cite{Aristov2009}, in the case of a two-lead
junction, the contributions neglected in the above-mentioned partial
summation are subleading and may be shown to vanish when the fixed points
are approached. 

Generally, the transport behavior of one-dimensional systems at low
energy/temperature is dominated by only a few fixed points (FPs) of the RG
flow. The most common and intuitively plausible FPs are those with quantized
conduction values, $G=0,1$, in units of the conductance quantum $%
G_{0}=e^{2}/h$ (we consider systems of spin-less fermions). There may,
however, appear additional FPs associated with noninteger conductance
values, usually in the case of attractive interaction. A particularly
well-studied case is that of a symmetric three-lead junction with broken
chiral symmetry, as induced by a magnetic flux threading the junction. This
latter problem has been studied by the method of bosonization in \cite%
{Chamon2003,Oshikawa2006}, and by the functional RG method in\cite%
{Barnabe2005}. These authors have identified a number of new FPs, for
attractive interaction. The discovery of these new FPs has sparked interest
in the problem of mapping out the complete fixed point structure of the
theory, even though most of the interesting new physics appears to be in the
physically less accessible domain of strongly attractive interactions. Only
the fully symmetric situation of identical wires ($g=g_{3}$, see below) has
been considered in \cite{Oshikawa2006,Barnabe2005}. As we have shown in \cite%
{Aristov2012a} for weak coupling and will show below for any coupling
strength, important new physics is missed by restricting the consideration
to the symmetric case. 

One result found by  \cite{Oshikawa2006} deserves special attention: it is
the claim that a new fixed point named $D$ (Dirichlet) exists for strongly
attractive interaction (Luttinger parameter $K>3$), which is interpreted as describing
multiparticle transport. As a consequence, the conductance values at this FP
are outside the domain of values allowed by a single particle $S$-matrix description.
Put differently, this requires the excitation of particle-hole pairs at the
junction, with say the particle moving into lead $2$, while the hole is
going into lead $3$. In our view,  at least at zero temperature, and in the
linear response regime, such processes are not occurring, because the phase
space for particle-hole excitations tends to zero in this limit. As we
will show below, we do not find a fixed point with the characteristics of 
$D$, but in contrast we find two new fixed points, not discussed in Refs.\ [\onlinecite{Oshikawa2006,Hou2012}]. 

In principle, the Y-junction set-up may be realized experimentally by a
one-dimensional tunneling tip contacting a quantum wire. A magnetic flux
threading the junction may be created by local magnetic moments at the
junction, although this has not been realized experimentally so far.

In this paper we report results of an RG treatment of electron transport in
the linear response regime through a junction of three spinless TLL wires
threaded by a magnetic flux. We employ a fermionic representation as
described in detail in\cite{Aristov2009}.  The method has formerly been used
in \cite{Das2004,Lal2002} in the case of Y-junctions. More recently this
approach has been used to derive the RG equations for the conductances of a
Y-junction connecting three TLL wires in the absence of magnetic flux, for
weak interaction \cite{Aristov2010}, and in a recent work for any strength
of interaction \cite{Aristov2011a}. In\cite{Aristov2010} it was found that
even in weak coupling, but beyond lowest order, interesting new structures
appear. Even a weak higher order contribution may change the RG-flow in a
dramatic way. In the present paper we describe a similar effect: an
asymmetry of the three wires of a chiral Y-junction to the effect that in
the tip wire the interaction strength $g_{3}$ is different from that in the
main wire, $g$, allows to access certain regions in interaction parameter
space where a whole line of fixed points rather than a single fixed point is
stable. This happens for attractive interaction only ($g,g_{3}<0$) \ The
line of stable fixed points is connecting two fixed points at two special
manifolds of interaction values, (a) $g=0$; $g_{3}<0$, or (b) whenever the
condition $K_{3}^{-1}=2-(K+K^{-1})/2$,  $1< K < 2$ is met. 
Along these fixed point lines the FP
values of the conductances are continuously varying. Results on this new
aspect of transport through Y-junctions within weak coupling have
been reported in \cite{Aristov2012a}.  In the present work we extend our
theory into the strong coupling regime. We derive a general expression for
the scale-dependent contribution to the conductances, valid for a junction
with any number of leads and any values of interaction $g_{j}$ in lead $j$ .
This result allows to derive RG-equations for the set of independent
conductances in the given case. We go on to evaluate these general
expressions for the three-lead junction of the $Y$-type (interaction $g$ in
the main wire and $g_{3}$ in the tip wire) in the presence of magnetic flux.
Then we determine the six physically allowed FPs, and discuss their
stability as a function of position in coupling constant space. 

In order to compare our results with the recent findings of \cite{Hou2012} for a fully asymmetric junction,  we extend our analysis by allowing for full asymmetry of the S-matrix at the junction, while keeping equal interaction in the main wires. This minimal modification enables us to elucidate the differences between our conclusions and those in \cite{Hou2012}. We show that the set of RG equations for the most general asymmetric chiral case can be easily derived in our formalism, and allows us to discuss the robustness of our results obtained for the partially asymmetric case ($g_{1}=g_{2}$) with respect to further asymmetry. A full analysis of the case $g_{1} \neq g_{2}$ is, however, beyond the scope of the present study. 
 
\begin{figure}[tbp]
\includegraphics*[width=0.8\columnwidth]{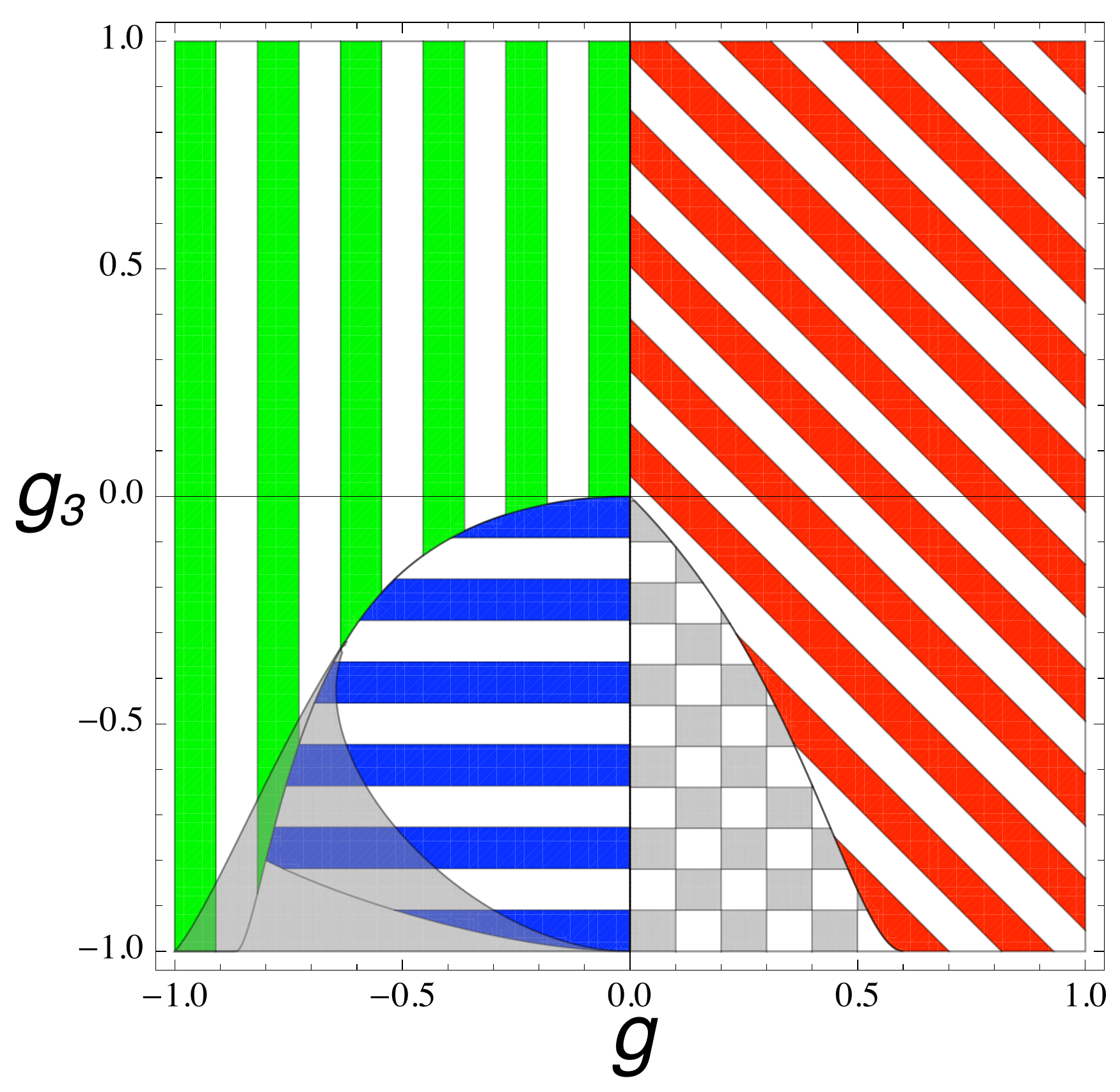}  
\caption{(Color online) The RG phase portrait, showing regions with different stable fixed points.  Diagonal red stripes correspond to stable $N$ point, vertical green stripes - to $A$ point, horizontal blue stripes - to $\chi^{\pm}$ points, gray shaded and checkerboard regions indicate stable $M$ point. }
\label{fig:Portrait2}
\end{figure}

Our findings for the 1-2-symmetric Y-junction are summarized in Fig. \ref{fig:Portrait2}, where the regions of stability of the four FPs $N$, $A$, $\chi^{\pm}$ and $M$ are displayed in the space of interaction constants $g$ and $g_{3}$. Fixed point $N$ corresponds to the complete separation of all three wires. The separation of the third wire from the perfectly conducting ``main wire'' 1-2 is described by FP $A$. The chiral FPs 
$\chi^{+}$, $\chi^{-}$ correspond  to maximum chirality. They describe a kind of Hall effect situation (depending on the orientation of the magnetic flux), where the in-current from wire $j$ flows into wire $j+1$ ($\chi^{+}$) or  $j-1$ ($\chi^{-}$). Fixed point $M$, finally, corresponds to the situation of maximum transparency. The stability of the $M$-point in the region to the right, $g>0$, (checkerboard pattern) is lost for infinitesimally small deviations from perfect 1-2 symmetry  of the junction, in favor of a new $A$-point ($A_{1}$ or $A_{2}$, meaning separation of wire 1 or 2 from the remaining ``main wire''). The region of stability of the $M$-point on the left, $g<0$, remains (see the detailed discussion in Section \ref{sec:anisotropy} and Fig.\ \ref{fig:PortraitK}). It is worth noting that in certain regions two FPs (e.g., $A$ and $M$, or $\chi^{\pm}$ and $M$) are stable and attract trajectories within their respective basins of attraction.  

\section{\label{sec:PTcond} Perturbation theory for conductances}

\subsection{The model}

We consider a model of interacting spinless fermions describing three
quantum wires connected at a single junction by tunneling amplitudes in the
presence of a magnetic flux piercing the junction. In the continuum limit,
linearizing the spectrum at the Fermi energy and including forward
scattering interaction of strength $g_{j}$ in wire $j$, we may write the TLL
Hamiltonian in the representation of incoming and outgoing waves in lead $j$
(fermion operators $\psi _{j,in}$, $\psi _{j,out}$) as

\begin{eqnarray}
\mathcal{H} &=&\int_{-\infty }^{0}dx[H_{j}^{0}+H_{j}^{int}]\,,  \notag \\
H^{0} &=&v_{F}\Psi _{in}^{\dagger }i\nabla \Psi _{in}-v_{F}\Psi
_{out}^{\dagger }i\nabla \Psi _{out}\,, \\
H^{int} &=&2\pi \{g[\widehat{\rho }_{1}\widehat{\widetilde{\rho }}_{1}+%
\widehat{\rho }_{2}\widehat{\widetilde{\rho }}_{2}]+g_{3}\widehat{\rho }_{3}%
\widehat{\widetilde{\rho }}_{3}\}\Theta (x;-L,-l) \,.  \notag
\end{eqnarray}

Here $\Psi _{in}=(\psi _{1,in},\psi _{2,in},\psi _{3,in})$ denotes a vector
operator of incoming fermions and the corresponding vector of outgoing
fermions is expressed through the $S$-matrix as $\Psi _{out}(x)=S\cdot \Psi
_{in}(-x)$ . In the chiral representation we are using positions on the
negative (positive) semi-axis corresponding to incoming (outgoing) waves. We
consider quantum wires of finite length $L$, contacted by reservoirs. The
transition from wire to reservoir is assumed to be adiabatic (i.e. produces
no additional potential scattering). The junction is assumed to have
microscopic extension $l$ of the order of the Fermi wave length.\ Inside the
junction interaction effects are neglected. This is expressed by the window
function $\Theta (x;-L,-l)=1$, if $-L<x<-l$, and zero otherwise. The regions $x<-L
$ are thus regarded as reservoirs or leads labeled $j=1,2,3$. We denote the
interaction constants $g_{1}=g_{2}=g$ from now on, and put the Fermi
velocity $v_{F}=1$. The various incoming and outgoing channels are
illustrated in Fig.\  1 elsewhere. \cite{Aristov2011a,Aristov2012a}  
The interaction term of the Hamiltonian is expressed in
terms of density operators $\widehat{\rho }_{j,in}=\Psi ^{+}\rho _{j}\Psi =%
\widehat{\rho }_{j}$, and $\widehat{\rho }_{j,out}=\Psi ^{+}\widetilde{\rho }%
_{j}\Psi =\widehat{\widetilde{\rho }}_{j}$, where $\widetilde{\rho }%
_{j}=S^{+}\cdot \rho _{j}\cdot S$ and the density matrices are given by $%
(\rho _{j})_{\alpha \beta }=\delta _{\alpha \beta }\delta _{\alpha j}$ and $(%
\widetilde{\rho }_{j})_{\alpha \beta }=S_{\alpha j}^{+}S_{j\beta }$. The $S$%
-matrix characterizes the scattering at the junction and (up to irrelevant
phase factors) has the structure (see  \cite{Aristov2012a})

\begin{equation}
S=\left( 
\begin{array}{ccc}
r_{1} & t_{12} & t_{13} \\ 
t_{21} & r_{1} & t_{23} \\ 
t_{31} & t_{32} & r_{2}%
\end{array}%
\right) \label{Smatrix}
\end{equation}

A convenient representation of 3$\times $3-matrices is in terms of Gell-Mann
matrices $\lambda _{j}$, $j=0,1,\ldots ,8$, the generators of $SU(3)$ (see%
\cite{Aristov2011a}). Notice that the interaction operator only involves $%
\lambda _{3}$ and $\lambda _{8}$ (besides the unit operator $(\lambda
_{0})_{\alpha \beta }=\sqrt{2/3}\,\delta _{\alpha \beta }$). We note $%
\mbox{Tr}(\lambda _{j})=0$, $\mbox{Tr}(\lambda _{j}\lambda _{k})=2\delta _{jk}$,  $%
j=1,\ldots ,8$, and $[\lambda _{3},\lambda _{8}]=0$. We introduce a compact
notation by defining a three-component vector $\overline{\mathbf{\lambda }}%
=(\lambda _{3},\lambda _{8},\lambda _{0})$, in terms of which the densities
may be expressed as $\rho _{j}=\sqrt{1/2}\sum_{\mu }R_{j\mu }\overline{%
\lambda }_{\mu }$, where the $3\times 3$ matrix $\mathbf{R}$ is defined as

\begin{equation}
\mathbf{R}=\left( 
\begin{array}{ccc}
\frac{1}{\sqrt{2}} & \frac{1}{\sqrt{6}} & \frac{1}{\sqrt{3}} \\ 
-\frac{1}{\sqrt{2}} & \frac{1}{\sqrt{6}} & \frac{1}{\sqrt{3}} \\ 
0 & -\sqrt{\frac{2}{3}} & \frac{1}{\sqrt{3}}%
\end{array}%
\right) 
\end{equation}%
and has the properties $\mathbf{R}^{-1}=\mathbf{R}^{T}$, $det\,\mathbf{R}=1$. 
The outgoing amplitudes are expressed in the analogous form with 
$\lambda_{j}$ replaced by $\widetilde{\lambda }_{j}=S^{+}\cdot \lambda _{j}\cdot S$. 
With the aid of the $\lambda _{j}$ the $S$-matrix may be parametrized by
eight angular variables (see \cite{Aristov2011} and Appendix \ref{sec:Smatrix}). 
For the case under
consideration only three of these, $\theta ,\psi ,\xi $, are relevant: $%
S=e^{i\lambda _{2}\xi /2}e^{i\lambda _{3}(\pi -\psi )}e^{i\lambda _{5}\theta} 
e^{i\lambda _{2} (\pi -\xi) /2}$. The corresponding elements of the $S$%
-matrix are given by 

\begin{equation}
\begin{aligned}
r_{1} &=\frac{1}{2}(\cos \theta +e^{i\psi })\sin \xi \,, \quad 
r_{2}=\cos \theta  \,,   \\
t_{13} &=t_{32}=i\cos \frac{\xi }{2}\sin \theta  \,, \quad 
t_{23}=t_{31}=i\sin \frac{\xi }{2}\sin \theta \,, 
\\
t_{12} &=\cos \theta \cos ^{2}\frac{\xi }{2}-e^{i\psi }\sin ^{2}\frac{\xi }{%
2}\,, \quad t_{21}=t_{12}|_{\xi \rightarrow \pi -\xi }  \,. 
\end{aligned}
\label{Smatrix:coef}
\end{equation}%
As will be shown below, the S-matrix and therefore the angular variables $%
\theta ,\psi ,\xi $, will be renormalized by the interaction.\bigskip 

\subsection{Linear response conductances}

In the linear response regime, the conductances
$$C_{jk}=\int_{-\infty }^{-L}dy\,\langle (\widehat{\rho }_{j}(-L)-\widehat{%
\widetilde{\rho }}_{j}(L))\widehat{\rho }_{k}(y)\rangle _{\omega \rightarrow
0}$$
relate the current $I_{j}$ in lead $j$ (flowing towards the junction) to the
electrical potential $V_{k}$ in lead $k$, $I_{j}=C_{jk}V_{k}$. Here the
combination $(\widehat{\rho }_{j}(-L)-\widehat{\widetilde{\rho }}_{j}(L))$
means that the current is measured at $x=-L$, and the integration over $y$
for the incoming densities is over the length of the ideal leads, to which
the electric potential $V_{k}$ is applied. Denoting the frequency of the applied external field by $\omega$, we consider the d.c. limit $\omega L\rightarrow 0$, 
when the value of $L$ does not play a role. 
\cite{Aristov2009}  The quantum averaging $\left\langle
\ldots\right\rangle $ involves taking the trace over the matrix products. The
conductance matrix $\mathbf{C}$ has only three independent elements, $G_{aa}=%
\frac{1}{2}(1-a)$, $G_{bb}=\frac{2}{3}(1-b)\,$, and $G_{ab}=c/\sqrt{3}\,$,
relating the currents $I_{a}=\frac{1}{2}(I_{1}-I_{2})$, $I_{b}=\frac{1}{3}%
(I_{1}+I_{2}-2I_{3})$ to the bias voltages $V_{a}=(V_{1}-V_{2})$, $V_{b}=%
\frac{1}{2}(V_{1}+V_{2}-2V_{3})$.  In compact notation we may define a $%
3\times 3$ matrix of conductances

\begin{equation}
\mathbf{G} =\left( 
\begin{array}{ccc}
G_{aa} & G_{ab} & 0 \\ 
-G_{ab} & G_{bb} & 0 \\ 
0 & 0 & 0%
\end{array}%
\right) =\left( 
\begin{array}{ccc}
\frac{1}{2}(1-a) & -\frac{c}{\sqrt{3}} & 0 \\ 
\frac{c}{\sqrt{3}} & \frac{2}{3}(1-b) & 0 \\ 
0 & 0 & 0%
\end{array}%
\right) 
\end{equation}%
The connection between matrices  $\mathbf{C}$ and  $\mathbf{G}$ comes from the observation, that 
 the nonzero elements of the matrix $\mathbf{C}^{R}=\mathbf{R}%
^{T}\mathbf{CR}$ are essentially the reduced conductances: $%
C_{11}^{R}=2G_{aa}\,$, $C_{22}^{R}=\frac{3}{2}G_{bb}\,$, $%
C_{12}^{R}=-C_{21}^{R}=\sqrt{3}G_{ab}\,$, where the numerical factors arise
due to the physically motivated asymmetric definitions of the currents and
voltages. It may be shown that the parametrization of the conductance tensor
in terms of $a,b,c$ follows quite naturally, by observing that the initial
conductances following from the Kubo formula are given by $C_{jk}=\delta
_{jk}-\mbox{Tr}(\widetilde{\rho }_{j}^{\;r}\rho _{k})=\delta _{jk}-|S_{jk}^{\;r}|^{2}$
where the superscript $r$ denotes that the quantity is fully renormalized by
interactions \cite{Aristov2011a}. By expressing 
$Y_{jk}=\mbox{Tr}(\widetilde{\rho }_{j}^{\;r}\rho _{k})$ in terms of 
$Y_{\mu \nu }^{R}=\frac{1}{2}\mbox{Tr}(\widetilde{\lambda }_{\mu }^{r}\lambda _{\nu })$ as 
$\mathbf{Y}=(\mathbf{R\cdot Y}^{R}\cdot \mathbf{R}^{T}\mathbf{)}$, we find by
comparison with the conductance matrix, $\mathbf{C}^{R}=\mathbf{1-Y}^{R}$, that 
$\mathbf{Y}^{R}$ has nonzero elements given by the conductance parameters
introduced above:

\begin{equation}
\mathbf{Y}^{R}=
\begin{pmatrix} 
a & c & 0 \\ 
-c & b & 0 \\ 
0 & 0 & 1%
\end{pmatrix} 
\label{def:abc}
\end{equation}%
From now on we will drop the superscript $r$ , with the understanding that
all quantities are renormalized. Therefore, we may use $\mathbf{Y}^{R}$ to
represent the conductances in the renormalization group analysis below.

\begin{figure}[tb]
\includegraphics*[width=0.7\columnwidth]{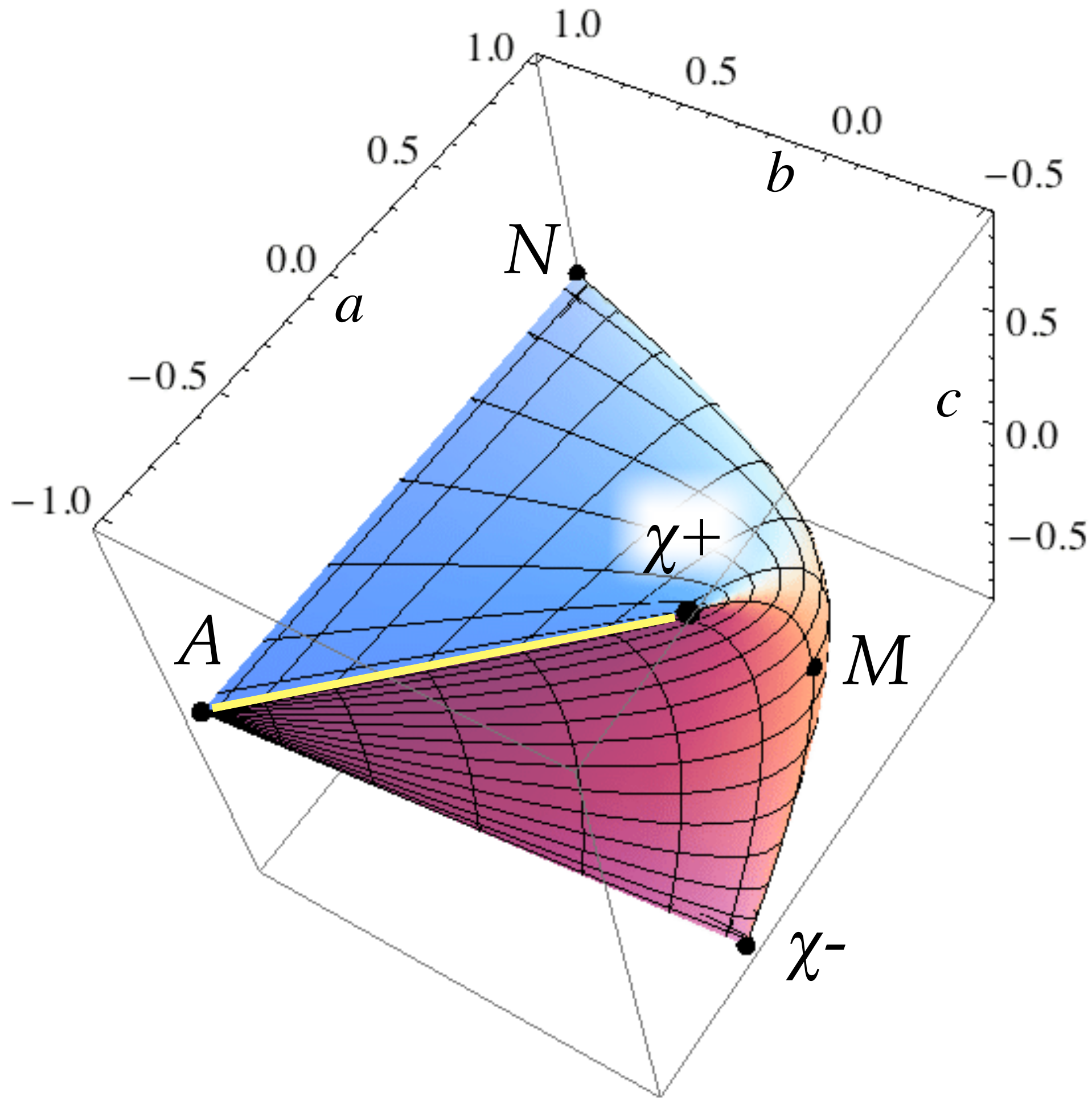}
\caption{\label{fig:body} (Color online) Allowed values of conductances, $a,b,c$, lie inside
the depicted body, $B$. The location of the RG fixed points $N$, $A$, $\protect\chi%
^{\pm}$, $M$ is also shown. The line of fixed points connecting FPs $A$ and $%
\protect\chi^{+}$ is indicated in yellow. }
\end{figure}

By substituting the $S$-matrix in the form (\ref{Smatrix}), (\ref{Smatrix:coef}) into the
definition of $\mathbf{Y}^{R}$, we find the relation of the $a$, $b$, $c$ to
the Euler angle variables: $a=-\frac{1}{2}(\cos ^{2}\theta +1)\cos ^{2}\xi
+\cos \theta \cos \psi \sin ^{2}\xi $, $b=\frac{1}{2}(3\cos ^{2}\theta -1)$, 
$c=-\frac{\sqrt{3}}{2}\sin ^{2}\theta \cos \xi $. We find therefore that $
a,b,c$ are confined within the region $a\in [ -1,1]$, $b\in [-1/2,1]$, $c\in [ -\sqrt{3}/2,\sqrt{3}/2]$. The physically allowed
points in $a$-$b$-$c$-space, satisfying the conditions $|\cos \theta |<1$, 
$|\cos \psi| <1$ and $|\cos \xi| <1$, lie inside a body labelled $B$ as shown in
Fig.\ \ref{fig:body}. We may trace the above restriction on the allowed
values of the conductance back to our consideration of energy conserving
scattering. In the linear response regime and at zero temperature only lead
electrons at the Fermi level will be transported through the interacting
wire, and particle-hole excitations are excluded. The restriction of the set
of conductances to the domain of allowed values has important consequences
for the stability of certain fixed points of the RG flow, as will be
discussed below. We mention now and will show later that all fixed points
are located on the surface of the body $B$ in Fig.\ \ref{fig:body}. 

\begin{figure}[tbp]
\includegraphics*[width=0.8\columnwidth]{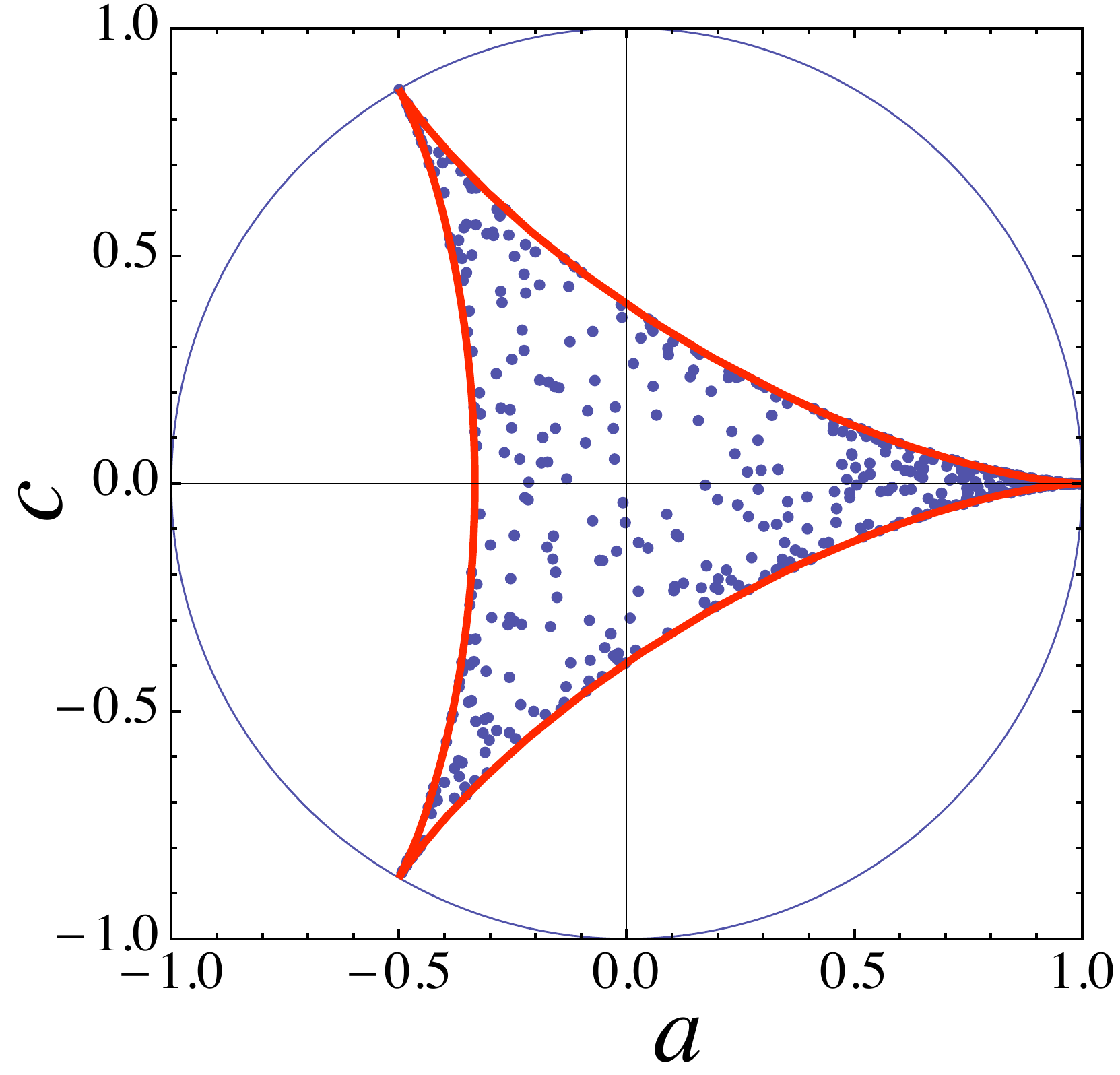}  
\caption{\label{fig:deltoid}  (Color online) 
Allowed values of conductance $a$, $c$ for the isotropic case.  
The S-matrices  (\protect\ref{Smatrix-tightbinding})
 characterized by parameters $\Gamma$, $\phi$ 
lead to two quantities $a$ and $c$, as  described by (\protect\ref{ac:tight}). We took 
500 pairs $\Gamma$, $\phi$ at random, these are blue points on the plot.
As discussed in Appendix \ref{sec:GammaPhi}, the domain $\Delta$ is bounded by a deltoid curve. 
}  \end{figure}

For the
isotropic case discussed below we have $a=b$.  One possible parametrization 
of the S-matrix for this case is given by Eq.\ (\ref{Smatrix-tightbinding}) 
The corresponding subspace of
allowed conductance values is the twodimensional domain $\Delta$ enclosed by a
deltoid curve, see Eq.\ (\ref{deltoid}) and Fig.\  \ref{fig:deltoid}. 

\section{Renormalization group equations}

The renormalization of the conductances by the interaction is determined by
first calculating the correction terms in each order of perturbation theory.
We are in particular interested in the scale-dependent contributions
proportional to\ $\Lambda =\ln (L/l)$, where $L$ and $l$ are two lengths,
characterizing the interaction region in the wires (see above). In lowest
order in the interaction the scale dependent contribution to the
conductances is given by \cite{Aristov2012a}

\begin{equation}
C_{jk}=C_{jk}\big\rvert_{\mathbf{g}=0}
+\tfrac{1}{2}\sum\limits_{l,m} \mbox{Tr}\left[\widehat{W}_{jk}\widehat{%
W}_{lm} \right] g_{ml}\Lambda \,,
\label{RG1order}
\end{equation}%
where  $C_{jk}\rvert_{\mathbf{g}=0}=\delta _{jk}-Y_{jk}$, the $\widehat{W}_{jk}= [\rho_{j}, \widetilde\rho_{k}]$ are a set of
nine $3\times 3$ matrices (products of $\widehat{W}$'s are matrix products), 
$g_{ml}=g_{m}\delta _{ml}$ is the matrix of interaction constants and the
trace operation $\mbox{Tr}$ is defined with respect to the $3\times 3$ matrix space
of $\widehat{W}$'s. As argued in \cite{Aristov2009}
the leading terms in each order of perturbation theory are certain diagrams
of the ladder type, which may be summed up analytically. The result is a
renormalized interaction matrix $\mathbf{g}^{ladder}$ replacing the bare
interaction matrix $\mathbf{g}$ \ in Eq.(\ref{RG1order}). The components of $\mathbf{g}%
^{ladder}$ are obtained from the following matrix equation (see appendix \ref{sec:App1})

\begin{equation}
\mathbf{g}^{ladder}=2(\mathbf{Q-Y})^{-1}\,.
\label{eq:gladder}
\end{equation}%
We observe that the effective interaction is found to depend on the
conductance components $C_{jk}=\delta _{jk}-Y_{jk}$ in a highly nonlinear
way. The matrix $\mathbf{Q}$ characterizes the interaction strength and
depends on the Luttinger parameters $K_{j}=[(1-g_{j})/(1+g_{j})]^{1/2}$ as

\begin{equation}
Q_{jk}=q_{j}\delta _{jk},\ \ \ q_{j}=(1+K_{j})/(1-K_{j})\,.
\end{equation}
(Notice that $|q_{j}|>1$ for any interaction strength.) 
In order to remove the redundancy in the conductance matrix $\mathbf{C}$ we
now multiply with $\mathbf{R}^{T}$ from the left and $\mathbf{R}$ from the
right to get the components of $\mathbf{Y}^{R}$ in the form

\begin{equation}
Y_{jk}^{R}=  Y_{jk}^{R} \big\rvert_{\mathbf{g}=0}-\frac{1}{2}\sum\limits_{l,m} 
\mbox{Tr}\left[\widehat{W}_{jk}^{R}\widehat{W}_{lm}^{R}\right]
g_{ml}^{ladder,R}\Lambda \,.
\end{equation}

Differentiating these results with respect to $\Lambda $ (and then putting $%
\Lambda =0$) we find the RG equations up to infinite order in the interaction

\begin{equation}
\frac{d}{d\Lambda }Y_{jk}^{R}=-\frac{1}{2}\sum \limits_{l,m}
\mbox{Tr}\left[\widehat{W}_{jk}^{R}\widehat{W}_{lm}^{R}\right] g_{ml}^{ladder,R} \,.
\label{eq:RGgeneral}
\end{equation}%
Here the $\widehat{W}_{jk}^{R}=\{\mathbf{R}^{T}\cdot \widehat{\mathbf{W}}%
\cdot \mathbf{R}\}_{jk}$ are a set of nine $3\times 3$ matrices and the
trace operation $\mbox{Tr}$ is defined with respect to that matrix space, whereas $%
g_{ml}^{ladder,R}=\{\mathbf{R}^{T}\cdot \mathbf{g}^{ladder}\cdot \mathbf{R}%
\}_{ml}= 2 \{(\mathbf{Q}^{R}-\mathbf{Y}^{R})^{-1}\}_{ml}$ \ and $\mathbf{Q}^{R},%
\mathbf{Y}^{R}$ are scalars with respect to this space. The nine matrices $%
\widehat{W}_{jk}^{R}$ are best evaluated with the aid of computer algebra,
inserting the $S$-matrix in terms of the quantities $a,b,c$ as given above (see also  Appendix \ref{sec:appGenAsym}).
As a result one finds the following set of RG-equations

\begin{widetext}
\begin{equation}
\begin{aligned}
\frac{da}{d\Lambda } &=D^{-1}%
\{a^{2}(3b-1-2Q_{1})+a(3c^{2}+q(1-b))+(Q_{1}-b)(1+b)+c^{2}(2+q)\}\,,  
\\
\frac{db}{d\Lambda } &=D^{-1}%
\{q(1-b)(1+2b)+a(b-1)(1+3b-Q_{1})+c^{2}(2+3b+Q_{1})\}\,, 
\\
\frac{dc}{d\Lambda } &=-cD^{-1}\{1+q+Q_{1}-3c^{2}+2b(q-1)+a(2Q_{1}-3b-2)\}%
\,.  
\end{aligned}
\label{RGfull} 
\end{equation}
\end{widetext}
where $D=(a-q)(b-Q_{1})+c^{2}$, and 
$$Q_{1}=\frac{3qq_{3}-q-2q_{3}}{2q+q_{3}-3}.$$ 
The RG-equations describe the flow of the conductances upon 
increasing $\Lambda$ until a stable fixed point is reached. The fixed points are found
by putting the right hand sides (the $\beta $-functions) simultaneously equal
to zero. The approach towards any given fixed point is characterized by
power laws. The fixed point pattern and the power law exponents depend on
the interaction strength. We note that the $\beta $-functions (the right
hand sides of Eqs.\ (\ref{RGfull}) ) are highly nonlinear functions of $a, b, c$ and of $%
q, q_{3}$, giving rise to a rich manifold of fixed points and RG flow
patterns. 

\section{\label{sec:FPs}Fixed points of RG flow}

\subsection{Chiral isotropic case}

\begin{figure*}[tbp]
\includegraphics*[width=0.8\columnwidth]{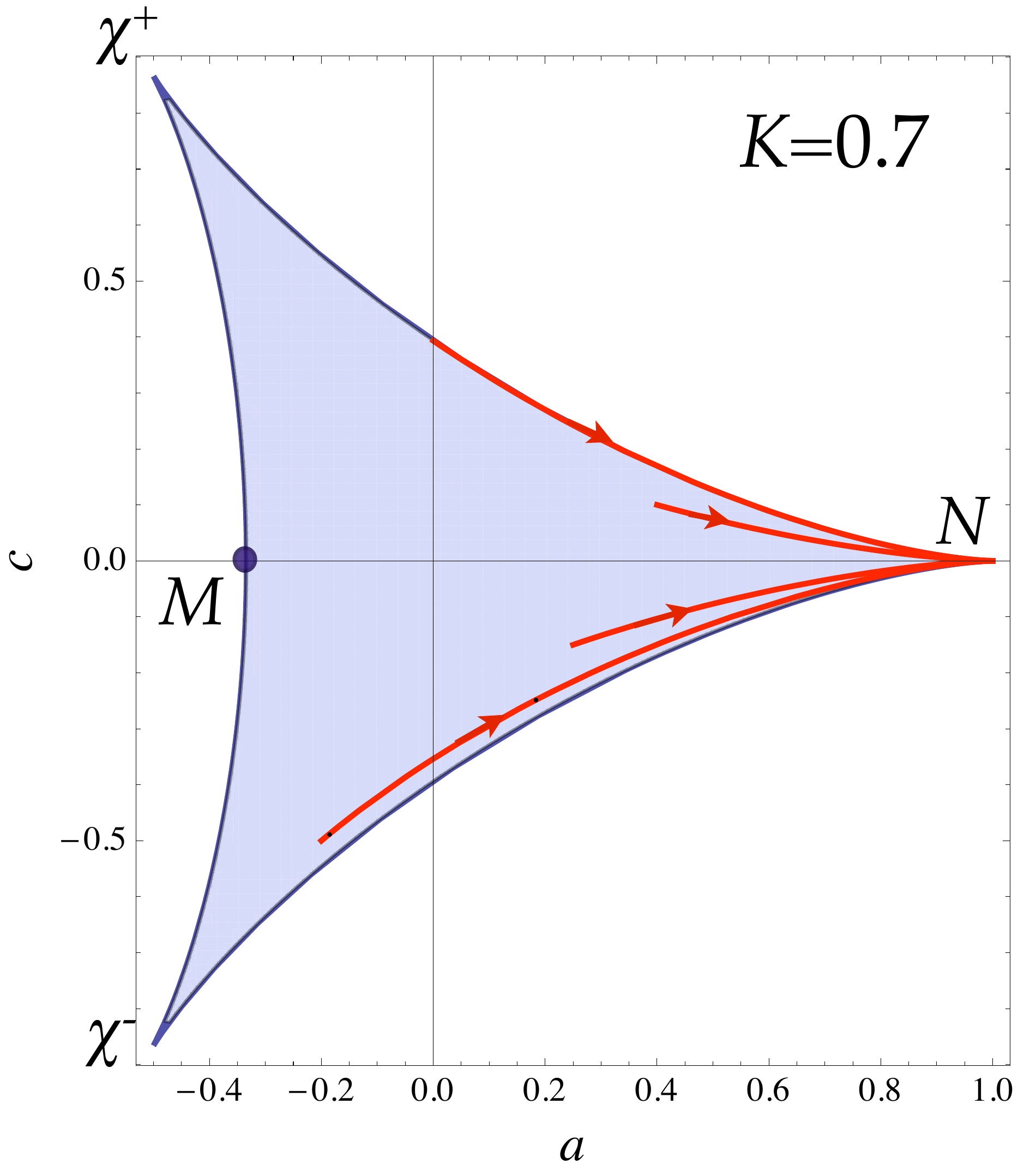} %
\includegraphics[width=0.8\columnwidth]{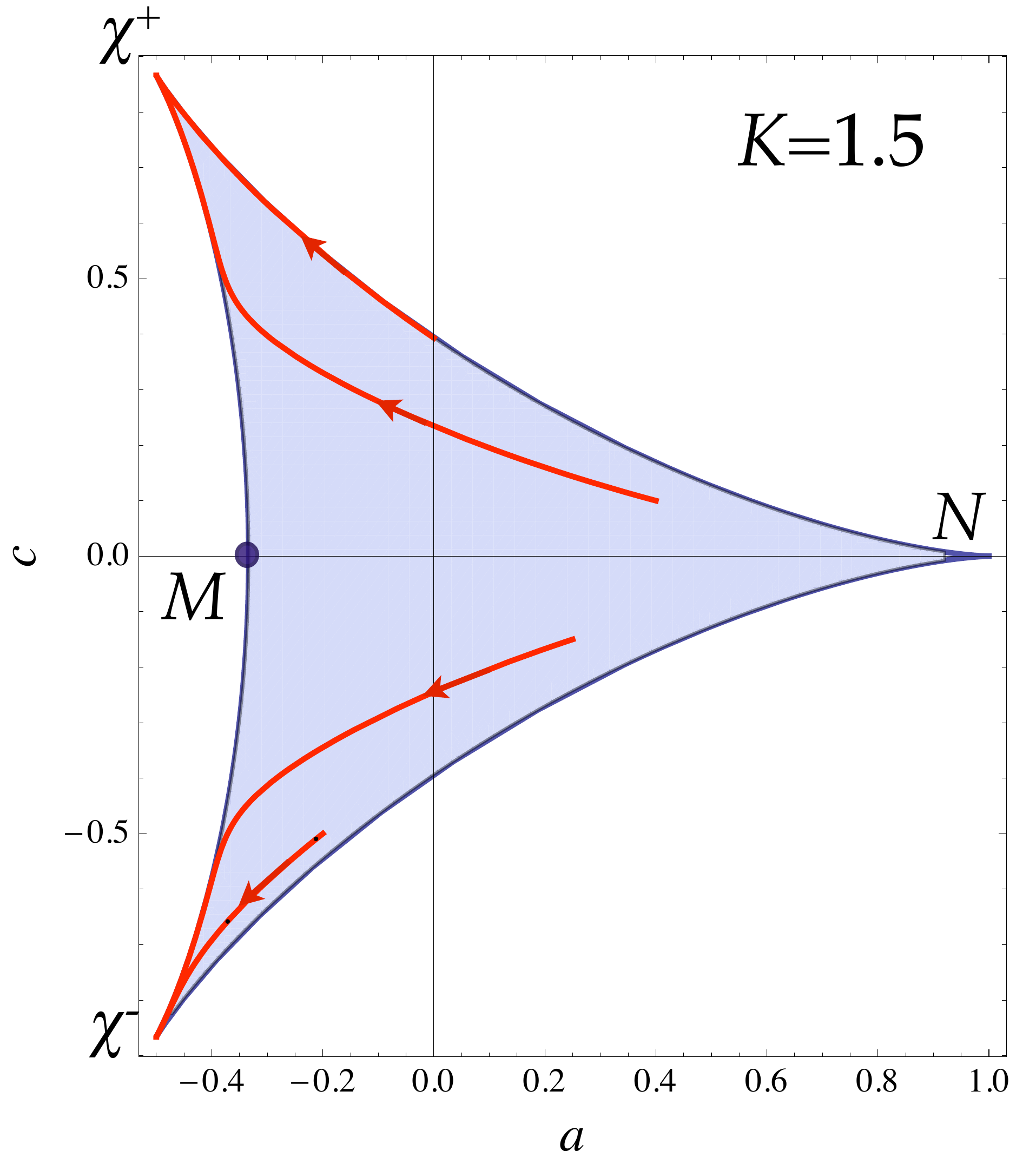} 
\includegraphics*[width=0.8\columnwidth]{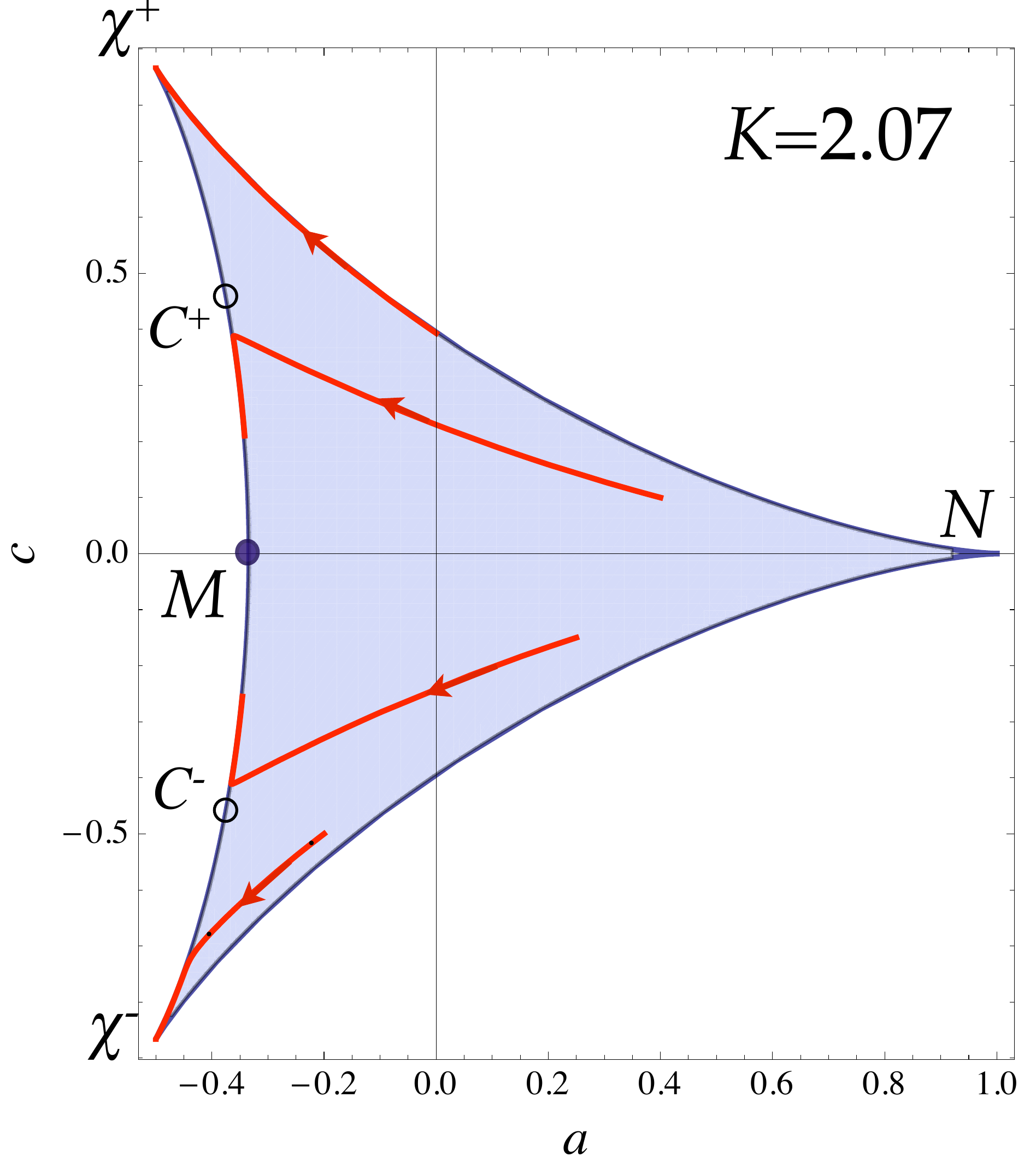} %
\includegraphics*[width=0.8\columnwidth]{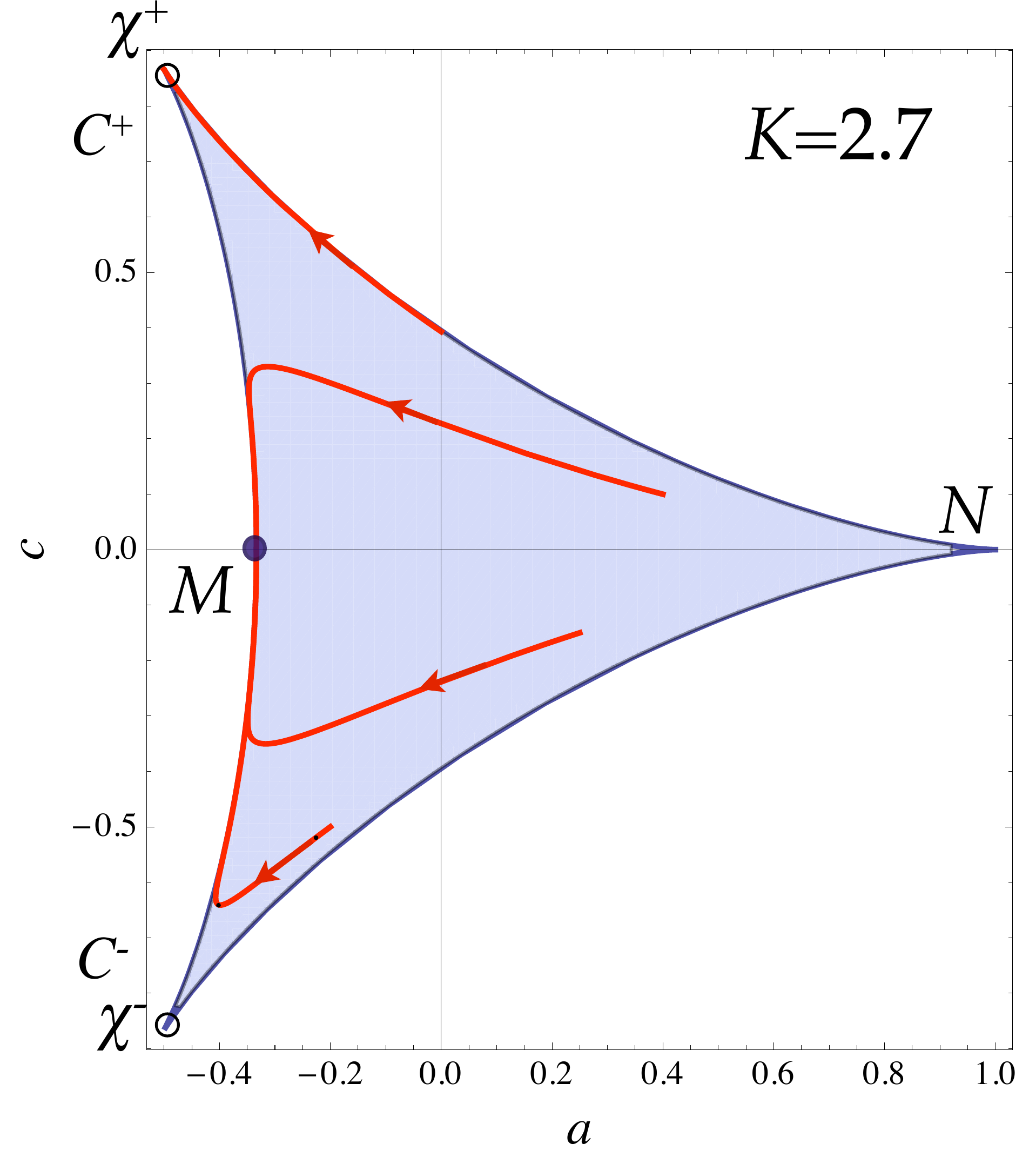} 
\includegraphics*[width=0.8\columnwidth]{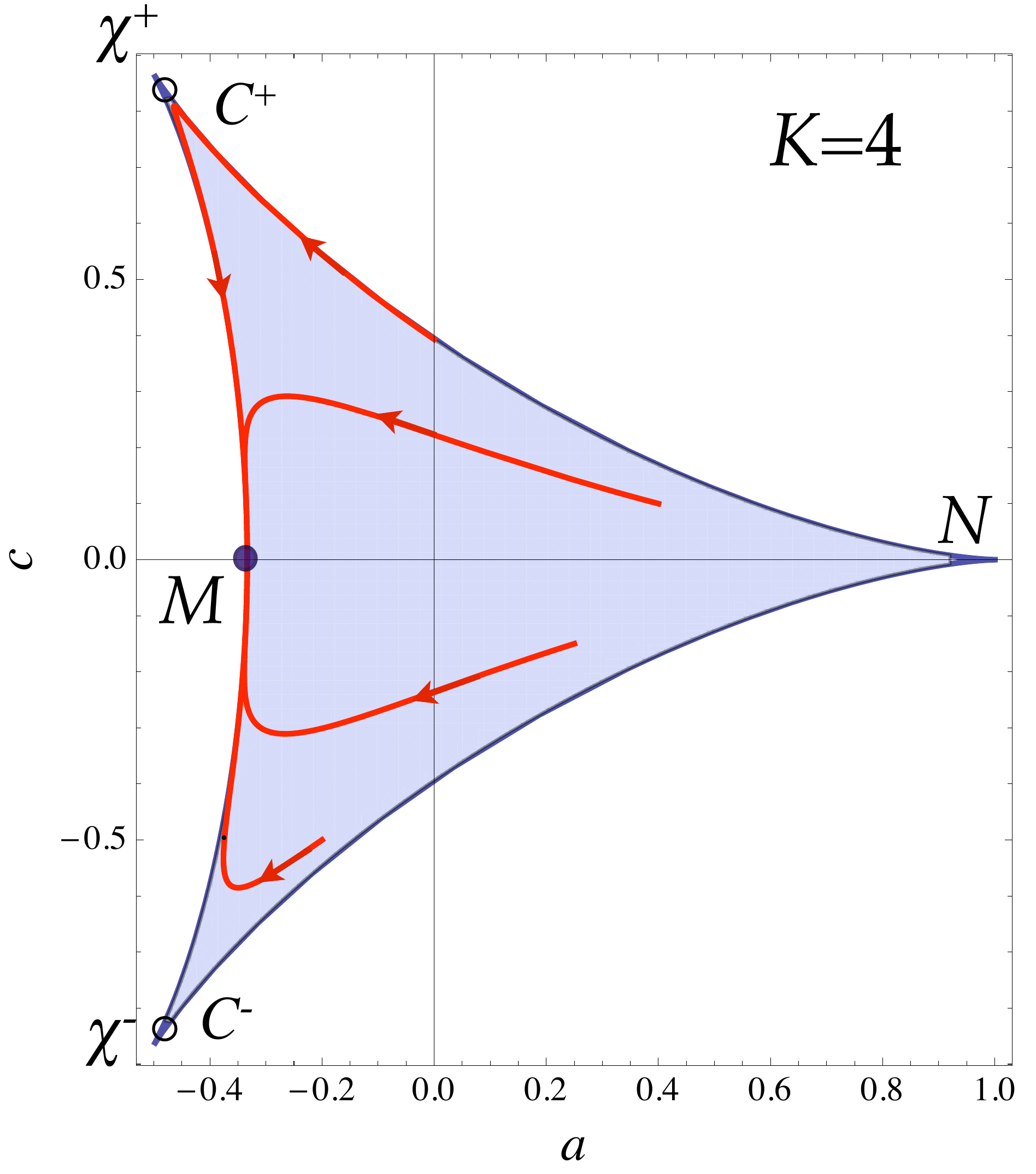} %
\includegraphics*[width=0.8\columnwidth]{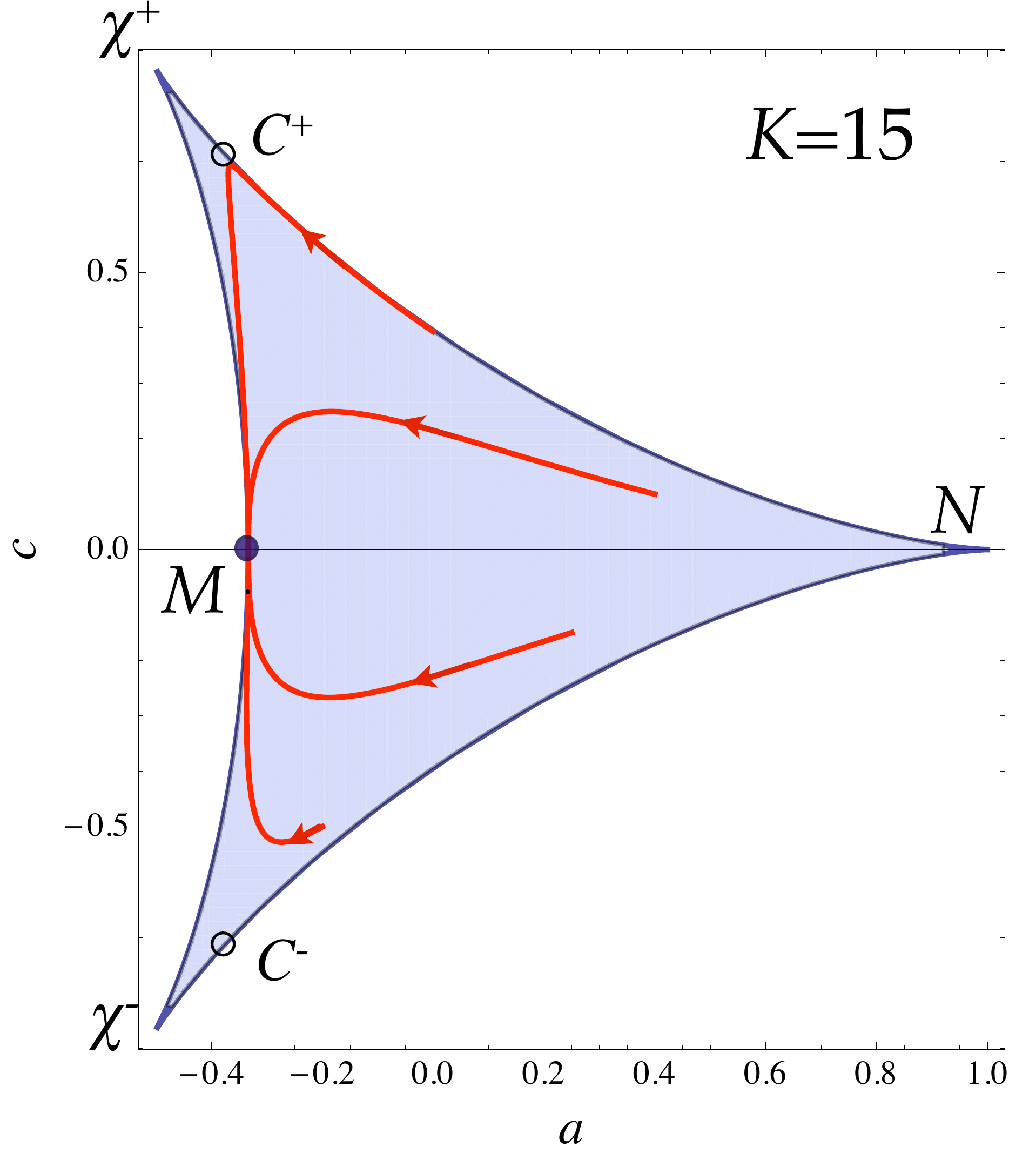}
\caption{(Color online) RG flows for the fully symmetric chiral case, for different values
of interaction. The four starting points are the same in all plots. The
non-universal positions of the chiral $C^{\pm }$ points are shown by open
circles. }
\label{fig:RGflows}
\end{figure*}

In the isotropic case, meaning (i) equal interaction strength in all wires,  $q=q_{3}$,  and (ii) isotropic tunneling amplitudes,  we have $Q_{1}=q$ and $a=b$. The RG
equations read

\begin{equation}\begin{aligned}
\tfrac{da}{d\Lambda } &=\beta
_{a}=D_{is}^{-1}\{(q-a)(1-a)(1+3a)+c^{2}(2+q+3a)\}\,,    \\
\tfrac{dc}{d\Lambda } &=\beta
_{c}=-cD_{is}^{-1}\{1+2q+4a(q-1)-3(a^{2}+c^{2})\}\,.   
\end{aligned}\end{equation} 
where $D_{is}=(a-q)^{2}+c^{2}$. By putting the $\beta $-functions equal to
zero we find four fixed points, labeled $N,M,\chi ^{\pm },C^{\pm }$ , at
positions

\begin{equation} \begin{aligned}
a_{N} &=1 , \qquad   c_{N}=0,   \\
a_{M} &= -1/3 , \quad    c_{M}=0,   \\
a_{\chi ^{\pm }} &=-1/2, \quad c_{\chi ^{\pm }}=\pm \sqrt{3}/2, \\
a_{C^{\pm }} &=\frac{3-K^{2}}{3(K-1)^{2}}, \quad c_{C^{\pm }}=\pm 
\frac{2K\sqrt{K(K-2)}}{3(K-1)^{2}} \,.
 \end{aligned} 
 \label{isotropicFPs}
 \end{equation} 
The location of these points in the $c-a$-plane is shown in Fig.\ \ref{fig:RGflows}. We
observe that all the FPs lie on the boundary curve of the domain of allowed
conductances. We discard the unphysical solution $a=q$, $c=0$, because $|a|<1<|q|$.

Next we discuss the stability of these FPs. We start with the $N$-point,
located at a cusp of the boundary curve $\Delta$, see Fig.\ 2. The cusp is
infinitely sharp, leaving only a single direction of approach from the
inside of $\Delta$ : $a=1-x$ , $c=0$ . Linearizing the RG equation for $a$ in the
small quantity $x$ we find

\begin{equation}
\frac{dx}{d\Lambda }=-2x(K^{-1}-1) \,. \notag
\end{equation}%
It follows that the $N$-point is stable for repulsive interaction, when $%
K^{-1}-1>0$ . 

We now turn to the $\chi ^{\pm }$-FPs, which are again located at a cusp of
Curve $\Delta$, Fig.\  \ref{fig:deltoid}. 
Therefore, again there is only one allowed direction for the
trajectory towards or away from $\chi ^{\pm }$ : $(a,c)=\frac{1}{2}(-1,\pm 
\sqrt{3})(1-x)$ , where $x$ is found to obey

\begin{equation}
\frac{dx}{d\Lambda }=-2x \left(\frac{4K}{3+K^{2}}-1\right) \,. \notag
\end{equation}%
We thus see that the $\chi ^{\pm }$-FPs are stable in the interaction regime 
$1<K<3$ .

The $M$-point is located at a flat section of curve $A$ and we therefore
have a two-dimensional manifold of possible trajectories leading to it.
Expanding the RG-equations in linear order around the $M$-point we find 
\begin{eqnarray}
\frac{da}{d\Lambda } &=&6\frac{1-K}{2+K}\left(a+\frac{1}{3} \right) 
,  \notag \\
\frac{dc}{d\Lambda } &=&-3\frac{(1-K)(2-K)}{(2+K)^{2}}c 
 \,.\notag
\end{eqnarray}%
It follows that the $M$-point is stable for $K>2$ . We note in passing that
the $M$-point cannot be discussed within the Abelian bosonization approach,
as the criterion $a_{M}^{2}+c_{M}^{2}=1$ is not satisfied.  \cite{Aristov2011}

Whereas the $N,M,\chi ^{\pm }$-FPs have been described before, we find in
addition a new pair of FPs  $C^{\pm }$ here. The location of these FPs is
not geometrical, but depends on the interaction. We first observe that the $%
C^{\pm }$-points are physically allowed (i.e. are located within the open
domain $\Delta$) only for sufficiently strong attractive interaction, $K>2$ (they
are always residing at the boundary curve $\Delta$). At $K=2$ they merge with the 
$M$-point. As $K$ increases the $C^{\pm }$-points start to move away from
the $M$-point in opposite direction until they end at the $\chi ^{\pm }$%
-points when $K=3\,$. Upon further increase of $K$ the $C^{\pm }$-points move
beyond the $\chi ^{\pm }$-points along the boundary of $\Delta$ , until they end
at $a=-\frac{1}{3}$ , $c=\pm \frac{2}{3}$ , in the limit $K\rightarrow
\infty $ . The stability analysis tells that the $C^{\pm }$-points are never
stable. 

As a result, we conclude, that for $K>3$ the only stable fixed point is the $%
M$ point, and not the unphysical $D$ point $a=-1$, $c=0$ suggested in  
\cite{Oshikawa2006}.  In Sec. 6.3 of that paper the authors say:
``While the $\chi _{\pm }$ fixed points are stable against a
small change in the flux, we do not know whether the $M$ fixed point has
such a stability. The simplest assumption is that it does not. In this case
the RG flows might go to the $\chi _{\pm }$ fixed points for any non-zero
flux $\phi $, starting from arbitrarily small $V$. Alternatively, it is
possible that the $M$ fixed point is stable against adding a small flux. In
that case, there would have to be additional unstable fixed points defining
the boundaries between the basins of attraction of the $M$, $\chi _{+}$ and $%
\chi _{-}$ fixed points. So again, an 'economy' principle suggests the
simple picture with only three stable fixed points'', for $1<K<3$. 
Our analysis shows, however, that this conjecture is not fully
correct, as the new unstable fixed points $C^{\pm }$ appear for $K>2$ ,
allowing the point $M$ to become stable. Examples of flow trajectories for
interaction strengths $K=0.7;1.5;2.07;2.7;4;15$ are shown in Fig.\ \ref{fig:RGflows}.

\subsection{Chiral anisotropic case}

The above RG equations, Eqs.\ (\ref{RGfull}),  describe the flow of $a,b,c$ towards the
stable fixed points (FPs) inside the body $B$ of allowed conductance values
(see Fig.\ \ref{fig:body}), depending on the initial conditions. The fixed points are
defined by the zeros of the $\beta$-functions on the r.h.s.\ of Eqs.\ (\ref{RGfull}).
Some of the fixed points are outside the allowed region $B$ in $a-b-c$-space
and will be discarded as unphysical. There are altogether six physical FPs
(see table \ref{tab:posFPs}), which may be classified into two groups, universal
(independent of the interaction) and non-universal FPs.

The three universal FPs are labeled $N,A,\chi ^{\pm}$. These FPs may be
thought to be of geometrical nature, as the corresponding $S$-matrix
elements are $1$ or $0$ . The FP $N$ describes the total separation of the
wires, meaning zero conductance in all components. By contrast, $A$
describes the separation of wire $3$ from the perfectly conducting wires 1--2. Finally, 
$\chi ^{\pm}$ are the FPs generated by chirality (left- or
right-handed).

The non-universal FPs are labeled $M,Q,C^{\pm}$. They only exist or are
physical in a limited region of interaction constant space, mainly for
attractive interaction. As shown in table \ref{tab:posFPs} the conductances at these FPs
depend in a complicated way on the interaction constants. 

The locations of the $M$ and $Q$ points have been given in \cite{Aristov2011a}. We reproduce the results here in the present notation. Both points can simultaneously exist 
only for attractive interaction, $g, g_{3} <0$. 
 Notice that for equal interactions $g=g_{3}$ we have $q=q_{3}=Q_{1}$, $\tau=|1+q|$ (see below) and we return to  (\ref{isotropicFPs}).


The FPs $M$ and $Q$ merge along a line in interaction constant space, defined by 
$\tau=0 $, where 
\begin{equation}
\tau = \sqrt{1+q^{2}+2Q_{1}}  \,.
\label{def:tau}
\end{equation} 
The end point of this line is at $K=2$, $K_{3}=4/3 $, and
was named ``tricritical'' in Ref.\  [\onlinecite{Aristov2011a}]. 

The location of the FPs $C^{\pm}$ is best
represented in terms of the angular variables $\theta ,\psi ,\xi $ as

\begin{equation}
\begin{aligned}
\cos \theta &=-\frac{1}{3}(2+Q_{1})\,, \qquad \cos \psi =1 \,,  \\
\cos \xi &=\pm \frac{\sqrt{3}}{5+Q_{1}}[3-(2q-Q_{1})(2+Q_{1})]^{1/2}  \,.
\end{aligned}
\end{equation}
The $C^{\pm }$ points lie on the surface of the body of physically allowed
conductances, since $\cos \psi =1$ corresponds to the outmost points of our
parametrization. The FPs $C^{\pm }$ are only in the physical domain if the
interaction constants satisfy the requirement $|\cos \theta |\leq 1$ ,
implying $-5\leq Q_{1}<1$, and $0\leq \cos ^{2}\xi \leq 1$ leading to the
bounds 
\begin{equation*}
0\leq 3-(2q-Q_{1})(Q_{1}+2)\leq \tfrac{1}{3}(5+Q_{1})^{2}
\end{equation*}%
The interval shrinks to zero when $Q_{1}\rightarrow -5$, and the above inequalities demand $q\rightarrow -3$,
which in turn gives $q_{3}\rightarrow -7$. Remarkably, these values again
correspond to $K=2$ and $K_{3}=4/3$, dubbed the ``tricritical'' point above. 

From the Table \ref{tab:posFPs} one may also verify that the points $C^{\pm }$ merge with 
$\chi^{\pm }$ at 
\begin{equation}
Q_{1}=-2\quad \Leftrightarrow \quad K^{-1}+2K_{3}^{-1}=1
\label{ChiCmerging}
\end{equation}%
implying, in the symmetric situation, the condition $K=K_{3}=3$ .

\begin{table*}
\caption{\label{tab:posFPs} The positions of the fixed points of the RG equations, (\ref{RGfull}). The quantity $\tau$ is defined in Eq.\ (\ref{def:tau}). 
 }
\begin{ruledtabular}
\begin{tabular}{c|ccc}
 FP & $a$ & $b$ & $c$ \\   \hline
 $N$ & 1 & 1  & 0 \\
$A$ &  $-1$ & 1 & 0 \\   
$\chi^{\pm}$ & $-\frac{1}{2}$ & $-\frac{1}{2}$ & $\mp \frac{\sqrt{3}}{2}$  \\
$M$ 
& $\frac{1}{3} \left(|q|- \tau\right) \mbox{sign}(q)$ &   $ \tfrac 16 ((|q|  - \tau)^{2} -3)  $ & 0 \\
 $Q$ 
& $\frac{1}{3} \left(|q| +\tau\right) \mbox{sign}(q)$ &   $ \tfrac 16 ((|q|  + \tau)^{2} -3) $ & 0 \\
$C^{\pm}$ & $\frac{1}{6} \left(2 q
   (Q_{1}+2)-Q_{1}^2+1\right)$ & $\frac{1}{6} (Q_{1}
   (Q_{1}+4)+1)$ & $\mp \frac{1}{6} (Q_{1}-1) \sqrt{3-(2 q
 -Q_{1}) (Q_{1}+2)}$ \\
\end{tabular}
\end{ruledtabular}
\end{table*}
 
\section{\label{sec:stability} Stability of fixed points}

The above set of equations (\ref{RGfull}) allows to perform a rather
straightforward analysis of the stability of FPs. Assuming that we have a
fixed point $(a_{0},b_{0},c_{0})$, we may expand the RG equations in terms
of the vector of small deviations 
$\mathbf{x} = (a-a_{0},b-b_{0},c-c_{0} ) 
= \lambda ( a_{1},  b_{1},  c_{1})$, 
to linear order in $\lambda \ll 1$:

\begin{equation}
\left[ \frac{d}{d\Lambda }-\mathbf{M} \right]\mathbf{x}=0
\end{equation}%
The stability of the fixed point is defined by the direction of the RG flow,
which is determined by the eigenvalues $\mu _{i}$ ($i=1,2,3$) of the matrix $\mathbf M$. All $\mu _{i}$ are positive for the fully unstable FP, all $\mu _{i}$
are negative for stable FPs and in the the saddle point case (some $\mu _{i}$
negative, some positive) the FP is again unstable.

Two remarks are in order here. First, the matrix $\mathbf M$ is in general not
symmetric, so that its eigenvectors are not orthogonal. Second, in the case
of asymmetric interaction ($q\neq Q_{1}$) the expressions for $\mathbf M$ are not
simple for the non-universal FPs, and should be analyzed numerically.
Simplifications occur at $q=q_{3}=Q_{1}$, as was already demonstrated for
the $M$ point in the non-chiral case in \cite{Aristov2011a}.

We mention that not all directions of the vector $(a_{1},b_{1},c_{1})$ are
permitted. This is because all FPs lie on the surface of the body of
physically allowed conductances. Certain displacements $\mathbf{x}$
would take the point $(a,b,c)$ outside of the body. If the FP lies on a
smooth part of the surface of the body, the requirement for $\mathbf{x}$ 
to be inside the body is not very restricting, because it only selects
half of all possible directions. But in our case the situation is more
complicated by the fact, that some FPs lie on ridges of the body. This
situation happens with the universal FPs and is discussed below.

\begin{figure}[tbp]
\includegraphics*[width=0.8\columnwidth]{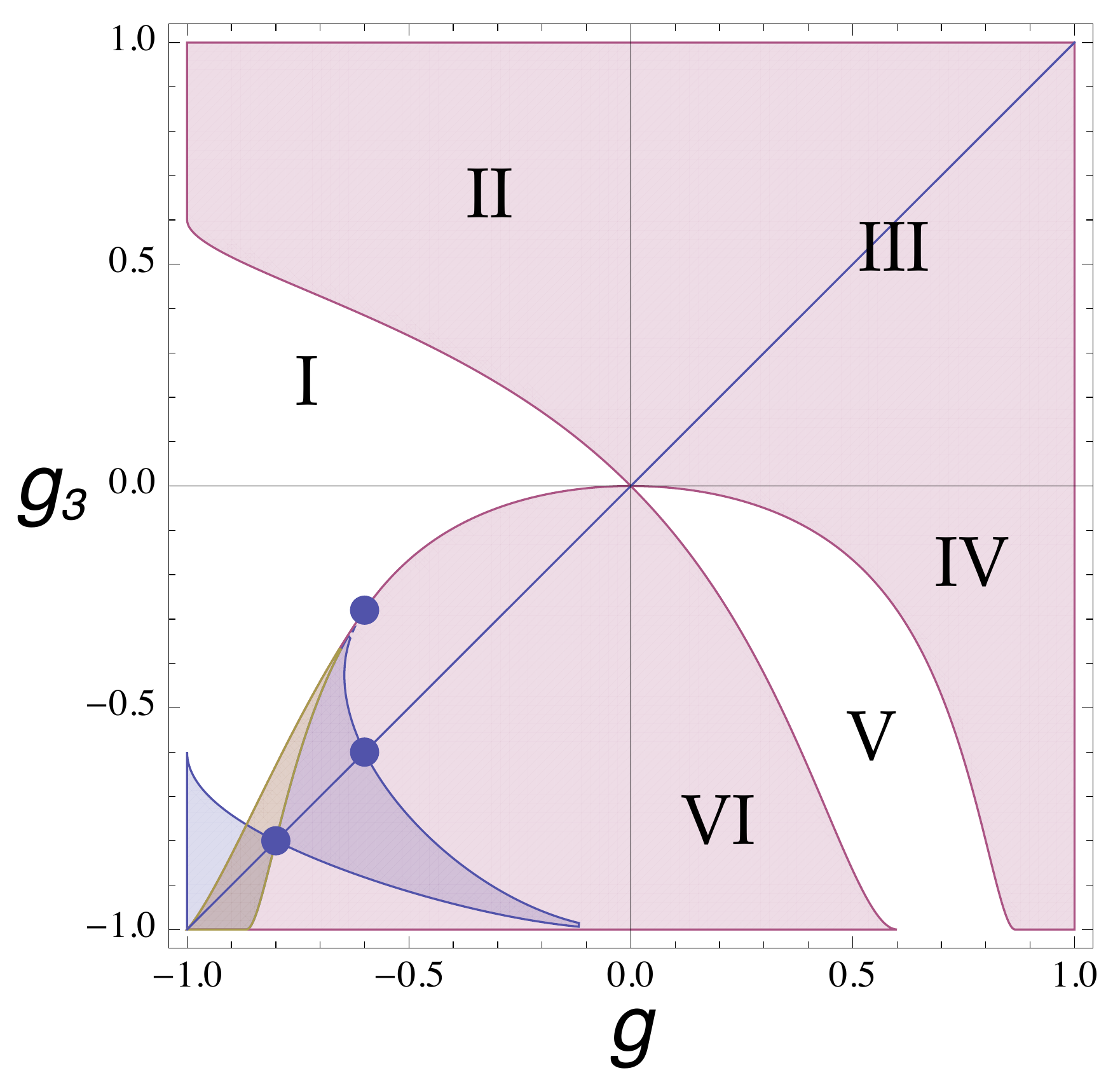} %
\includegraphics*[ width=0.8\columnwidth]{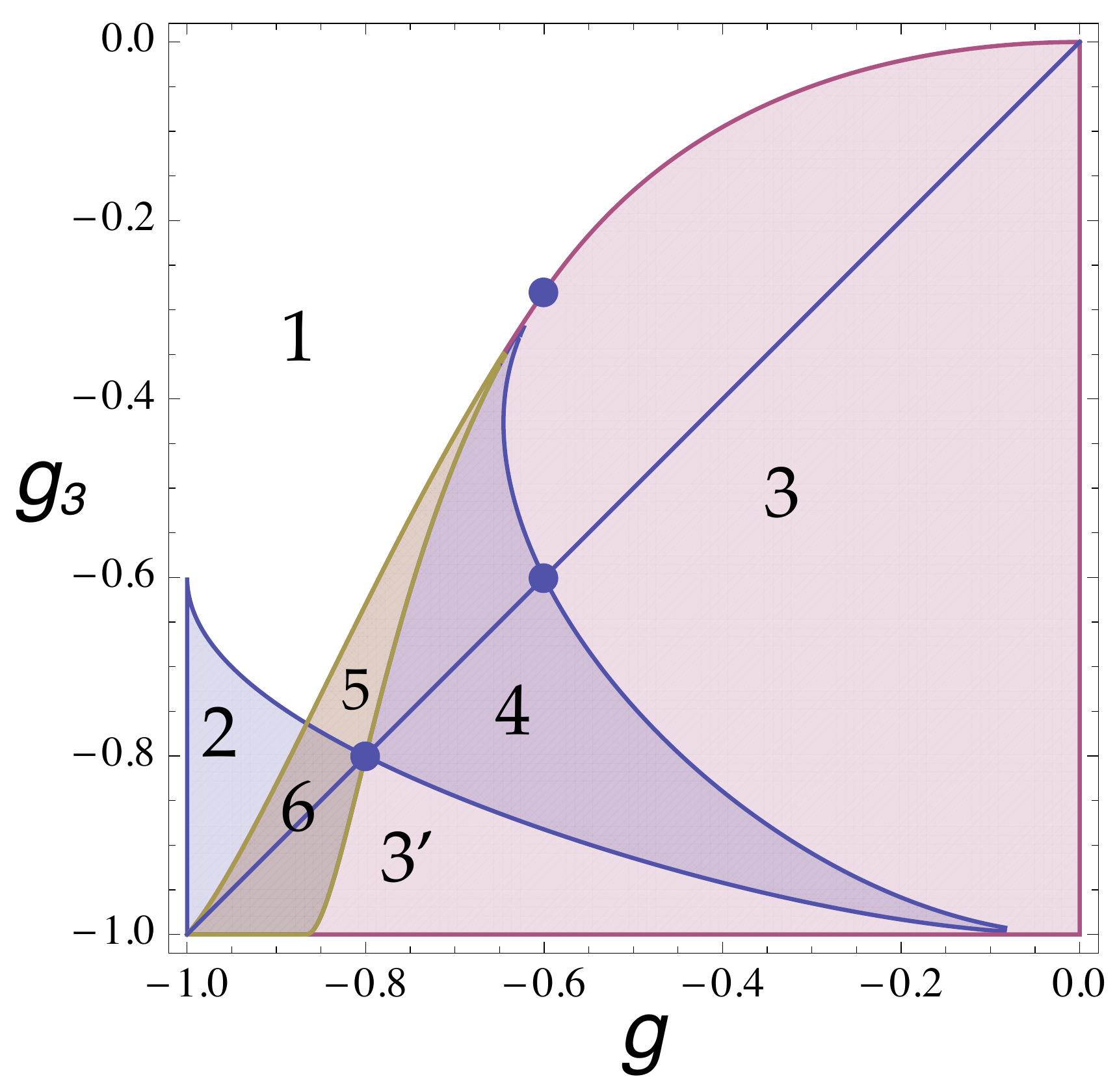}
\caption{(Color online) The RG phase portrait, showing regions with different set and
character of the FPs. The region of attractive interactions is shown in more
detail in the second panel. The numbers in the plot refer to its description
in Table \protect\ref{tab:PortraitSummary}. The three dots correspond to the special points 
$K=K_{3}=2$, $K=K_{3}=3$, and $K=2$, $K_{3}=4/3$, discussed in the text.}
\label{fig:Portrait}
\end{figure}

In the following we present a detailed stability analysis for the different fixed points.
The results are collected in Table \ref{tab:PortraitSummary} and in Fig.\ \ref{fig:Portrait}, 
showing the phase boundaries (i.e.\ the boundaries, separating different domains of stability) 
in the $g-g_{3}$-plane. 

\begin{table}[tbp]
\caption{Summary of the RG phase portrait, depicted in Fig.\ \protect\ref%
{fig:Portrait}. The existence and stability of the RG fixed points is shown
below. FPs $N$, $A$, $\protect\chi ^{\pm }$ always exist, and  the
stability is indicated by an ``s'' symbol. The
stable and unstable FPs $M$, $Q$, $C^{\pm }$ are denoted by
``s'' and ``u'', the ``-'' symbol is
used if such a point does not exist. }
\label{tab:PortraitSummary}%
\begin{ruledtabular}
\begin{tabular}{c|ccc|ccc}
& $N$ & $A$ & $\chi^{\pm}$ & $M$ & $Q$ & $C^{\pm}$\\   \hline
I &  u& s& u& -& -&-  \\ 
II & u & s & u& u& -&-  \\ 
III & s &u & u& u& -& - \\ 
IV & s &u & u& u& -&-  \\ 
V &  s &u & u&- & -& - \\ 
VI & u &u & u&s &- &-  \\ \hline
1 & u &s & u& -&- &  -\\ 
2 & u &s & u& -&- & u \\ 
3 & u &u &s & u& -&  -\\ 
4 & u &u &s &s &- &  u\\ 
5 & u & s&u & s& u&  -\\ 
6 & u & s & u&s & u& u \\
3' & u & u& u&s & -&  -
\end{tabular}
\end{ruledtabular}
\end{table}

A  corresponding figure for the non-chiral case ($c=0$) has been presented as Fig.\ 4 in our previous publication \cite{Aristov2011a}. By comparing the figures we observe that phases I--VI and 1,3',5  are characterized by FPs with $c=0$ and were already discussed in the non-chiral case. The new phases are labeled 2--4, 6, involving either the stable chiral FPs $\chi^{\pm}$ or the new unstable FPs, $C^{\pm}$. 

It should be emphasized, that the scaling exponents obtained by our method for the ``universal'' FPs  $N$, $A$, $\chi^{\pm}$ are in exact agreement with the results, found by boundary conformal field theory \cite{Hou2012}. 
Our results for the non-chiral FPs $N$, $A$ were reported earlier \cite{Aristov2011a}.  The particular expressions for the $\chi^{\pm}$ points below, Eq.\ (\ref{exp-chi}),  coincide with the exponents deduced from Table I of Ref.\  [\onlinecite{Hou2012}] for the scaling dimensions of the leading irrelevant boundary operators, $\Delta^{\chi}$. This becomes clear after the identification of our $\alpha _{\chi}$ with their $2 - 2\Delta^{\chi}$, and after putting their $g_{1}=g_{2}; g_{3}$ equal to our $K; K_{3}$.

\subsection{$N$ point}

In this case the matrix $\mathbf M$ is given by

\begin{eqnarray}
\mathbf{M}_{N} &=&f_{N}
\begin{bmatrix}
4(1-Q_{1}) & Q_{1}-q & 0 \\ 
0 & 4-3q-Q_{1} & 0 \\ 
0 & 0 & 3(2-q-Q_{1})
\end{bmatrix} \,,
\nonumber  \\
f_{N} &=&[(q-1)(Q_{1}-1)]^{-1}\,.
\end{eqnarray}%
Formally, there are three (left) eigenvectors $(1,-1/3,0)$, $(0,1,0)$, $%
(0,0,1)$, corresponding to eigenvalues $4/(1-q)$, $1/(1-q)+3/(1-Q_{1})$, $%
3/(1-q)+3/(1-Q_{1})$, respectively.

However, as can be seen in Fig.\ \ref{fig:deltoid}, any displacement with $%
c_{1}\neq 0$ is not allowed, as it would end outside the allowed domain.
Therefore, when discussing the RG flows starting inside ``the body'', we 
should discard any displacements along $c$.
We thus end with two exponents governing the RG
flow towards $N$

\begin{equation}
\begin{aligned}
\alpha _{N,1} &=4/(1-q)=-2(K^{-1}-1) \,, \\
\alpha _{N,2} &=\frac{1}{1-q}+\frac{3}{1-Q_{1}}=-(K^{-1}+K_{3}^{-1}-2) \,,
\end{aligned}
\label{exp-N}
\end{equation}%
which may be interpreted as the weak link exponents in the main wire and
between the main wire and the tip, respectively (eq. (46) in \cite%
{Aristov2011a}). These values are negative for $K,K_{3}<1$, so that the $N$
point is stable in the case of repulsive interaction in all wires.

\subsection{$A$ point}

The matrix $\mathbf{M}$ for the asymmetric $A$ point is given by

\begin{eqnarray}
\mathbf{M}_{A} &=&f_{A}
\begin{bmatrix}
4(Q_{1}-1) & Q_{1}+q & 0 \\ 
0 & Q_{1}-4-3q & 0 \\ 
0 & 0 & Q_{1}-4-3q%
\end{bmatrix}  \,, \nonumber 
\\
f_{A} &=&[(q+1)(Q_{1}-1)]^{-1}\,.
\end{eqnarray}
We have three (left) eigenvectors $(1,1/3,0)$, $(0,1,0)$, $(0,0,1)$,
corresponding to eigenvalues $4/(1+q)$, $1/(1+q)+3/(1-Q_{1})$, $%
1/(1+q)+3/(1-Q_{1})$, respectively. The last eigenvalue is doubly
degenerate, and therefore we again have only two exponents

\begin{eqnarray}
\alpha _{A,1} &=&4/(1+q)=2(1-K) \,, 
\label{scalingApoint}  \\
\alpha _{A,2} &=&\frac{1}{1+q}+\frac{3}{1-Q_{1}}=-(\tfrac{1}{2}%
(K^{-1}+K)+K_{3}^{-1}-2)  \,, \nonumber 
\end{eqnarray}
which may be interpreted as i) the weak impurity exponent in the main wire
and ii) the tunneling exponent in the main wire and the boundary exponent in
the tip, respectively (eq.\ (47) in \cite{Aristov2011a}). At the
\textquotedblleft tricritical\textquotedblright\ point, $K=2$, $K_{3}=4/3$,  the
second exponent vanishes. This is a manifestation of the fact, that the
tricritical point is the endpoint of the line of stable fixed points in the
interaction parameter space, see Fig.\  \ref{fig:Portrait}.

\subsection{$\protect\chi ^{\pm }$ point}

We consider only one of these chiral points, $\chi ^{-}$, for which we obtain

\begin{eqnarray}
\mathbf{M}_{\chi } &=&f_{\chi }
\begin{bmatrix} 4Q_{1}+3q+\frac{19}{2} & 2Q_{1}+q+\frac{3}{2} & -\sqrt{3}(2q+1) \\ 
3(Q_{1}+\frac{1}{2}) & Q_{1}+6q+\frac{19}{2} & -\sqrt{3}(2Q_{1}+1) \\ 
\sqrt{3}(2Q_{1}-\frac{1}{2}) & \sqrt{3}(2q-\frac{1}{2}) & 9
 \end{bmatrix}  \,, \nonumber 
\\
f_{\chi } &=&2[(2q+1)(2Q_{1}+1)+3]^{-1}\,.
\end{eqnarray}
The three left eigenvectors are $(-3,\frac{4Q_{1}-6q+5}{2Q_{1}+1},2\sqrt{3})$%
, $(\frac{3(Q_{1}-1)}{Q_{1}+3q+2},\frac{2Q_{1}+3q+1}{Q_{1}+3q+2},-\sqrt{3})$%
, $(\sqrt{3},\sqrt{3},-2)$, corresponding to the eigenvalues $2f_{\chi
}(3q-Q_{1}+4)$, $4f_{\chi }(Q_{1}+2)$, $3f_{\chi }(Q_{1}+q+4)$,
respectively. The last eigenvector points out of \textquotedblleft the
body\textquotedblright\ and should be discarded, similarly to the situation
at the $N$ point.

The other two physical exponents are

\begin{equation}
\begin{aligned}
\alpha _{\chi ,1} &=\frac{4(3q-Q_{1}+4)}{(2q+1)(2Q_{1}+1)+3} \,,
 \\ &= 2- \frac{8KK_{3}}{2K+K_{3}+ K^{2} K_{3}} \,,\\
\alpha _{\chi ,2} &=\frac{8(Q_{1}+2)}{(2q+1)(2Q_{1}+1)+3} \,,
\\ &=2 -  \frac{4K(K+K_{3}) }{2K+K_{3}+ K^{2} K_{3}}\,.
\end{aligned}
\label{exp-chi}
\end{equation}
For $K_{3}=K$ we have both exponents equal :  
\begin{equation*}
\alpha _{\chi ,1}=\alpha _{\chi ,2}=2-8K/{(3+K^{2})} \,.
\end{equation*}%

When the $C^{\pm }$ and $\chi ^{\pm }$ merge, which happens at $Q_{1}=-2$,
the second exponent $\alpha _{\chi ,2}$ in (\ref{exp-chi}) vanishes.

For $K_{3}\neq K$ the form of the eigenvectors is not transparent, and it
might be useful to restore the parametrization in terms of Euler angles. The
coordinates of the $\chi ^{\pm }$ points are then given by $\theta =\pi /2$, 
$\xi =0,\pi $, for arbitrary $\psi $. For $\chi ^{-}$, we put $\psi =0$ and
expand $\theta =\pi /2+\theta ^{\prime }$, $\xi =\pi +\xi ^{\prime }$, with $%
\theta ^{\prime },\xi ^{\prime }\ll 1$. Then,  linearizing the resulting RG equations, 
one finds a decoupled set of differential equations for 
$\theta ^{\prime },\xi ^{\prime }$

\begin{equation}
\begin{aligned}
\frac{d\theta ^{\prime }}{d\Lambda } &=\theta ^{\prime }\frac{2(3q-Q_{1}+4)%
}{(2q+1)(2Q_{1}+1)+3} \,, \\
\frac{d\xi ^{\prime }}{d\Lambda } &=\xi ^{\prime }\frac{4(Q_{1}+2)}{%
(2q+1)(2Q_{1}+1)+3}\,, 
\end{aligned}
\end{equation}
in correspondence with the above exponents for the conductance components.

\subsection{$M$, $Q$ and $C^{\pm}$ points}

As seen in Table \ref{tab:posFPs}, the position of these FPs is not universal, i.e. depends on the interaction. The linearization of the corresponding set of RG equations around these
points leads to very cumbersome expressions for the matrix $\mathbf{M}$, which
we do not show here.  Instead we list the three scaling exponents for the conductances.

For $M$ fixed point we have (for $g<0$): 
\begin{eqnarray}
\alpha _{M,1} &=&
-\frac{3 \tau  ((q+\tau)^{2} -9)}{2(q+\tau ) (2q-\tau)^2  }   \,, \nonumber \\
\alpha _{M,2} &=&   
-\frac{3  \left((q+\tau )^2+3\right)}{(q+\tau )(2 q-\tau ) } \,, \\
\alpha _{M,3} &=&  
-\frac{3 (q+\tau +3) ( q^{2}+2q-\tau^{2}+2\tau -3)}{2  (q+\tau )(2q-\tau)^2  } \,. \nonumber 
\end{eqnarray}
with the corresponding eigenvectors 
$(\frac{q ( (q + \tau)^{2}-9)}{3 + (5 q - 3 \tau) (q + \tau)}, -1, 0)$, 
$\left(q + \tau, -1,   0 \right)$, $(0, 0, 1)$ and $\tau$ defined in (\ref{def:tau}).
  
For $Q$ point we obtain (again, for $g<0$)
\begin{eqnarray}
\alpha _{Q,1} &=&
\frac{3 \tau  \left((q-\tau )^2-9\right)}{2 (q-\tau ) (2 q+\tau )^2}  \,, \nonumber \\
\alpha _{Q,2} &=&  
 -\frac{3 \left((q-\tau )^2+3\right)}{(q-\tau ) (2 q+\tau )} \,, \\
\alpha _{Q,3} &=&  
-\frac{3 (q-\tau +3) ( q^{2}+2q-\tau^{2}-2\tau -3)}{2 (q-\tau ) (2 q+\tau )^2}
 \,, \nonumber 
\end{eqnarray}
and the  eigenvectors are 
$(\frac{q ( (q - \tau)^{2}-9)}{3 + (5 q +3 \tau) (q - \tau)}, -1, 0)$, 
$(q - \tau, -1,   0)$, $(0, 0, 1)$, correspondingly.

 For the $C^{\pm}$ point we get 
 \begin{eqnarray}
\alpha _{C,1} &=&
\frac{12}{-3 q+Q_1+2}-\frac{12}{Q_1-1}-6  \,, \nonumber \\
\alpha _{C,2(3)} &=&   
\frac{3 \left(3 q-Q_1+4\right) \left(Q_1+5\right)}{2 \left(Q_1-1\right)  \left(-3
   q+Q_1+2\right)}
   \\ &\times& 
 \left[\frac{Q_1+1}{Q_1+5} \pm 
\frac13   \sqrt{1+
\frac{8 \left(Q_1+2\right) \left(3 q-Q_1-2\right)}{
   \left(Q_1-1\right) \left(3 q-Q_1+4\right)}
   }\right] \,,   \nonumber  
\end{eqnarray}
while the expressions for eigenvectors are too complicated to be presented here. 
One can check that at $Q_{1}=-2$ the exponent $\alpha _{C,2}$ vanishes, it corresponds to $C^{\pm}$ merging with $\chi^{\pm}$, as discussed after Eq.\  (\ref{exp-chi}).

\subsection{Lines of fixed points}

In our previous work \cite{Aristov2012a} on transport through a chiral $Y$%
-junction at weak coupling we found that at special lines in the  $g-g_{3}$%
-plane lines of fixed points may appear. We conjectured that this
may happen at any boundary in the  $g-g_{3}$-plane separating phases with
different stable FPs, which still exist upon approach to the boundary. A precondition is that the two different stable fixed points should exist as separate fixed points at the respective boundary in the  $g-g_{3}$-plane, i.e. the FPs should not have merged or otherwise disappeared upon approach to the boundary. By inspecting the many phase boundaries in Fig.4 in that respect we find that only at two phase boundaries, (1) separating phases $3$ and $VI$, at $K=1$; $K_{3}>1$ (i.e.\  $g=0$, $g_{3}<0$) and (2) separating phases $1$ and $3$, at $ \tfrac12 (K+K^{-1})+K_{3}^{-1}-2=0$, $1<K<2$ a line of fixed points emerges. Examples of trajectories in that case, demonstrating the existence of a line of fixed points, have been given in our previous work for the weak coupling regime \cite{Aristov2012a}.

\section{\label{sec:anisotropy} fully anisotropic Y-junction}
In the previous sections we presented a rather detailed analysis of the $1-2$ - symmetric Y-junction in the presence of magnetic flux. We now consider the effects of the absence of symmetry. As in previous cases, one might expect that some of the stable fixed points found above will become unstable if the additional freedom of the RG-flow in the absence of the symmetry restriction is admitted. One may distinguish two ways in which the symmetry with respect to wires 1-2 may be broken. The first is provided by different properties of the wires $1,2$, in particular different interaction strengths $g_{1} \neq g_{2}$. A second symmetry characterizes the junction, i.e. the way in which the third wire is attached. For instance, if the third wire is  non-perpendicular to the main wire, one may expect some asymmetry in hopping to wire 1 and to wire 2. The resulting asymmetry in the S-matrix  and the conductance matrix may grow in the process of renormalization. Such a scenario is implied in the work 
\cite{Hou2012}. In order to compare our results with those in \cite{Hou2012},  we now  discuss a possible anisotropy in the conductance tensor. 

In the following we confine ourselves for simplicity to the asymmetry of the second type, keeping in mind that the completely general case is fully covered by the general form of the RG-equations. 
The relevant formulas are listed in the Appendices \ref{sec:Smatrix}, \ref{sec:appGenAsym}. The S-matrix is now characterized by four Euler angles, and instead of Eq.\ (\ref{def:abc}) we have four conductance parameters
\begin{equation}
\mathbf{Y}^{R}=
\begin{pmatrix} 
a & c & 0 \\ 
\bar c & b & 0 \\ 
0 & 0 & 1%
\end{pmatrix} 
\label{def:abcd}
\end{equation}%
with $\bar c\neq -c$. In the asymmetric non-chiral situation we have $\bar c =c$, see  \cite{Aristov2011}. 
Using the general RG equation (\ref{eq:RGgeneral}) and formulas in Appendix \ref{sec:appGenAsym}, we obtain after some calculation: 

\begin{widetext}
\begin{equation}
\begin{aligned}
\frac{da}{d\Lambda } &=D_{1}^{-1}%
\{a^2  (3 b-1-2 Q_1 )+a (-3 c \bar c+q(1-b)) + (Q_{1}-b)(1+b)
 +c^2-c \bar c q+\bar c^2 \}\,,  
\\
\frac{db}{d\Lambda } &=D_{1}^{-1}%
\{q(1-b)(1+2b)+a(b-1)(1+3b-Q_{1})-c \bar c(2+3b+Q_{1})
\}\,, 
\\
\frac{dc}{d\Lambda } &=D_{1}^{-1}\{ c \left( a( 3 b-2 Q_1+2) -q (2 b+1)  \right)+\bar c \left(-2 b-3
   c^2+Q_1+1\right) \}
 \,, 
\\
\frac{d\bar c}{d\Lambda } &=D_{1}^{-1}\{ \bar c
   \left( a  (3 b- 2 Q_1+ 2 ) - q (2 b +1) \right) +c \left(-2 b-3 \bar c^2+Q_1+1\right) \}  
\,.  
\end{aligned}
\label{RGfullAsym} 
\end{equation}
\end{widetext}
with $D_{1}=(a-q)(b-Q_{1})-c \bar c $. In the limit $\bar c=- c$, we return back to (\ref{RGfull}).  

The investigation of the whole phase portrait defined by the set of equations, (\ref{RGfullAsym}), is beyond the scope of the present study. For our purposes it suffices to check the stability of the already discussed FPs in the slightly asymmetric situation, which can be done as follows.  Our previous analysis was confined to the surface $  c+\bar c=0$ in the space of conductance components $a,b,c,\bar c$.  We now allow for a small asymmetry of the Y-junction, $|c+\bar c  |\ll 1$ and expand the RG equations (\ref{RGfullAsym}) to first order in $h = c+\bar c$. As a result, we recover the three previously found equations  (\ref{RGfull}) and the fourth equation takes the form: 

\begin{equation}
\frac{dh}{d\Lambda }  =  h  D^{-1}
 (1-q +Q_{1} + 3 c^2 -2 b (q+1)- a  (2 Q_1- 3 b-2 ) )
 \label{RGeq:h}
\end{equation}
A negative (positive) coefficient of $h$ on the right-hand side of the last equation corresponds to stability (instability) of the previously determined FPs with respect to the asymmetric distortion of the S-matrix.

It is clear, that if the FP does not exist in some region of interaction parameter space, or is unstable, then the additional check of the stability provided by (\ref{RGeq:h}) is not required. Only otherwise stable FPs should be subjected to this additional test. 
 
Performing this analysis, we find that the FPs collected in Table  \ref{tab:PortraitSummary} are stable against asymmetry, with only one exception. Namely, the FP $M$ is unstable with respect to the asymmetry perturbation $h\neq0$ in region VI. In all other situations the character of the listed FPs is unchanged. Particularly, the exponents for $h$ around the points $N$ and $\chi^{\pm}$ are given by $\alpha_{N,2}$ and $\alpha_{\chi,2}$, Eq.\  (\ref{exp-N}) and (\ref{exp-chi}), respectively. The RG flow from the $M$ point in the region VI  should apparently lead to one of  two asymmetric FPs, different from $A$ and discussed in \cite{Aristov2011}. These points $A_{1,2}$ are defined by $a=1/2$, $b=-1/2$, $c=\bar c=\pm \sqrt{3}/2$ and correspond to the cases of either the first or the second wire detached from the remaining two wires. We will denote these points by  $\bar A$ below. In addition to these, we also may have the counterparts of the point $Q$ discussed above, at strong attraction. The unstable $Q$ point separates the $A$ and $M$ points, as discussed in   \cite{Aristov2011a}. Two new asymmetric $Q_{1,2}$ points appear for strong repulsion and their position is given by 
$a= \frac{1}{6} \left(2 q   \left(Q_1+2\right)+Q_1^2-1\right)$,  
$b= \frac{ 1}{6} \left(Q_1   \left(Q_1+4\right)+1\right)$, 
$c = \bar c = \pm \frac{1}{6} \left(Q_1-1\right) \sqrt{(2 q + Q_{1})   \left(Q_1+2\right) +3}$.

We notice also that the asymmetric FP $A$ cannot be affected by a finite value of $h$ without violating unitarity, meaning that the scaling behavior is determined by only two exponents, Eq.\ (\ref{scalingApoint}).  This situation is completely similar to what was discussed above, where the chiral perturbation was found impossible around the $N$ point, and one of the perturbations impossible around the chiral $\chi^{\pm}$ points. The corresponding boundary (deltoid) curve in the asymmetric non-chiral case, $c=\bar c$, was shown  in Fig.\ 1 of Ref.\  \cite{Aristov2011}. 

\begin{figure}[tbp]
\includegraphics*[width=0.95\columnwidth]{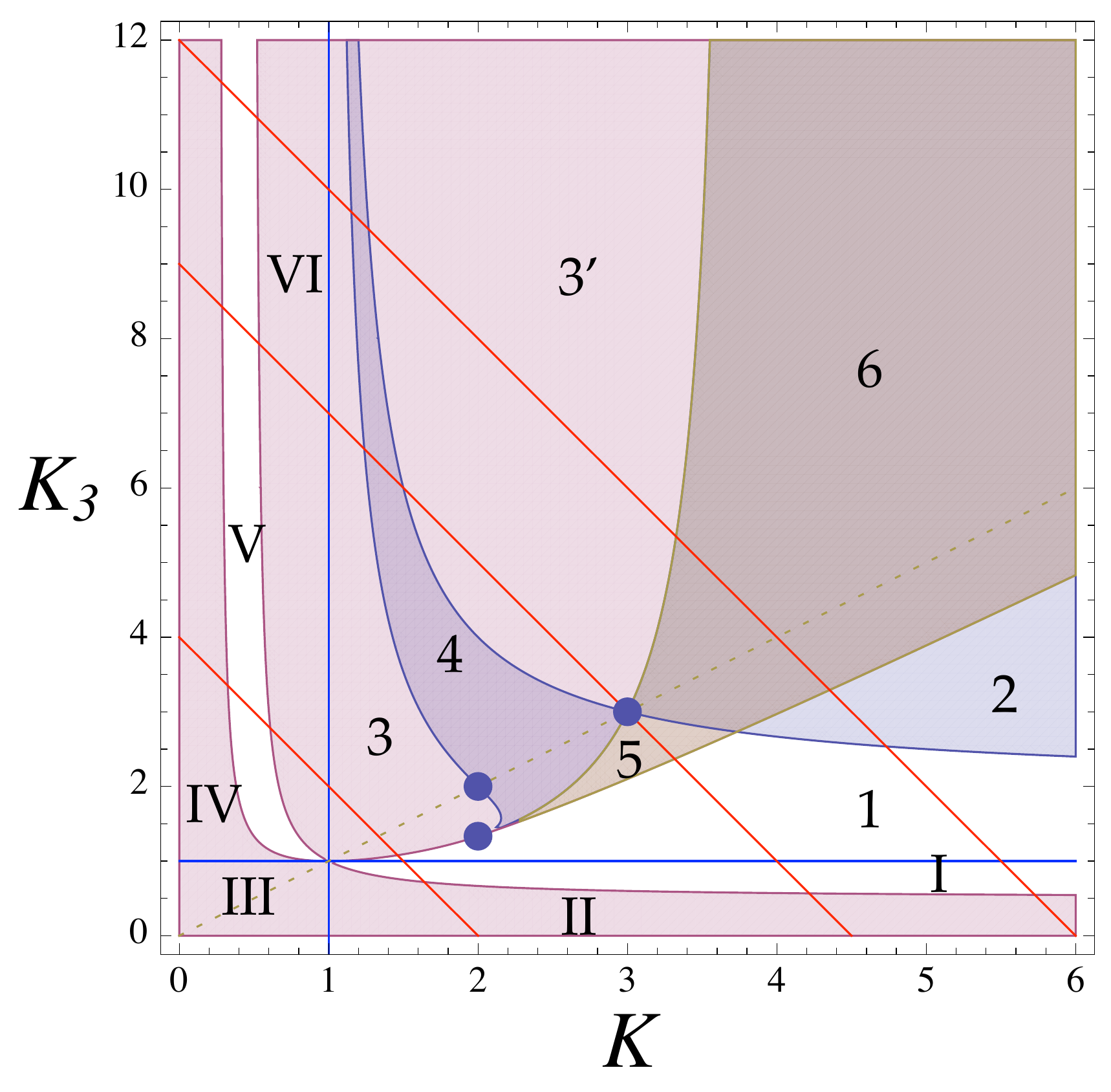}
\caption{\label{fig:PortraitK} (Color online)
The RG phase portrait, showing regions with different character of stable FPs.   The numbers in the plot refer to the description
in Table \protect\ref{tab:PortraitSummary}. The three dots mark the  points 
$K=K_{3}=2$, $K=K_{3}=3$, and $K=2$, $K_{3}=4/3$, discussed in the text. The three straight red lines correspond to $(2K+K_{3})/3 = 2,3,4$, respectively. The dashed line indicates the condition $K=K_{3}$. }
\end{figure}

Below we compare our results on the stability of different FPs with the results summarized in Fig.\ 5 of Ref.\ \cite{Hou2012}. In order to facilitate the comparison, we re-plot our Fig.\ \ref{fig:Portrait} in terms of the Luttinger parameters, $K$, $K_{3}$ and show the phase portrait in Fig.\ \ref{fig:PortraitK}. 

We observe full correspondence of our results with those shown in panels (a), (b), (c), (d) in Fig.\  5 of Ref.\ \cite{Hou2012} for relatively small interaction strength. However, we only partly agree with results obtained in \cite{Hou2012} for strong attraction, panels (e), (f), as discussed below. 

The three straight red lines in Fig.\ \ref{fig:PortraitK} designating the manifolds $(2K+K_{3})/3 = 2,3,4$, correspond to the vertical lines drawn through the top vertex of the triangles shown in panels (d), (e), (f) in Fig.\  5 of Ref.\ \cite{Hou2012}, respectively.  
 
We see that the first line, $(2K+K_{3})/3 = 2$, consecutively intersects the regions IV, V, VI, 3, 1, I, II, when one increases $K$ from $K=0$ to $K=2$. This sequence corresponds to the appearance of stable FPs in the following order:  $N$, $\bar A$, $\chi^{\pm}$, $A$, in full agreement with Fig.\ 5(d) of Ref.\ \cite{Hou2012}.  

The situation is more involved at stronger attraction. The second line,  $(2K+K_{3})/3 = 3$, consecutively intersects the regions IV, V, VI, 3, 4, 3', 5, 1, I, II,  in Fig.\ \ref{fig:PortraitK} upon the increase of $K$ from $K=0$ to $K=4.5$. From our Table \ref{tab:PortraitSummary} we see that this sequence corresponds to the FPs: $N$, $\bar A$, $\chi^{\pm}$, ($\chi^{\pm}$+$M$), $M$, ($M+A$), $A$.  It should be compared to the sequence
$N$, $\bar A$, $\chi^{\pm}$, $\bar A$,  $A$, shown in Fig.\  5(e) in \cite{Hou2012}. We notice two differences here: one is related to the fact that Hou \emph{et al.} \cite{Hou2012} do not include the $M$ point in their analysis, therefore they do not show the combination of several stable FPs, e.g. ($\chi^{\pm}$+$M$) in region 4, cf.\ Fig.\ \ref{fig:RGflows} at $K=2.07$ above. Another difference concerns the region $3'$ which appears, e.g.\ at $K=2$, $K_{3}=5$. Our analysis shows that, given the initial symmetry with respect to wires 1 and 2, the RG does not flow away from this symmetry, and the stable FP in this case is $M$.  This is in contradiction to the conclusion of the authors of  \cite{Hou2012}, who observe that the asymmetric points $\bar A$ are stable and hence any RG flow should end there. In fact, one can show from Eqs.\ (\ref{RGfullAsym}) that  the symmetric point $M$ is separated in the region 3'  from the asymmetric points  $ A_{1,2}$ by the unstable FPs $Q_{1,2}$, mentioned above. Note that the FPs $A_{1,2}$ and $Q_{1,2}$ appear only in the fully asymmetric case and therefore are not shown in Table \ref{tab:PortraitSummary}. 
 
\begin{figure}[tbp]
\includegraphics*[width=0.95\columnwidth]{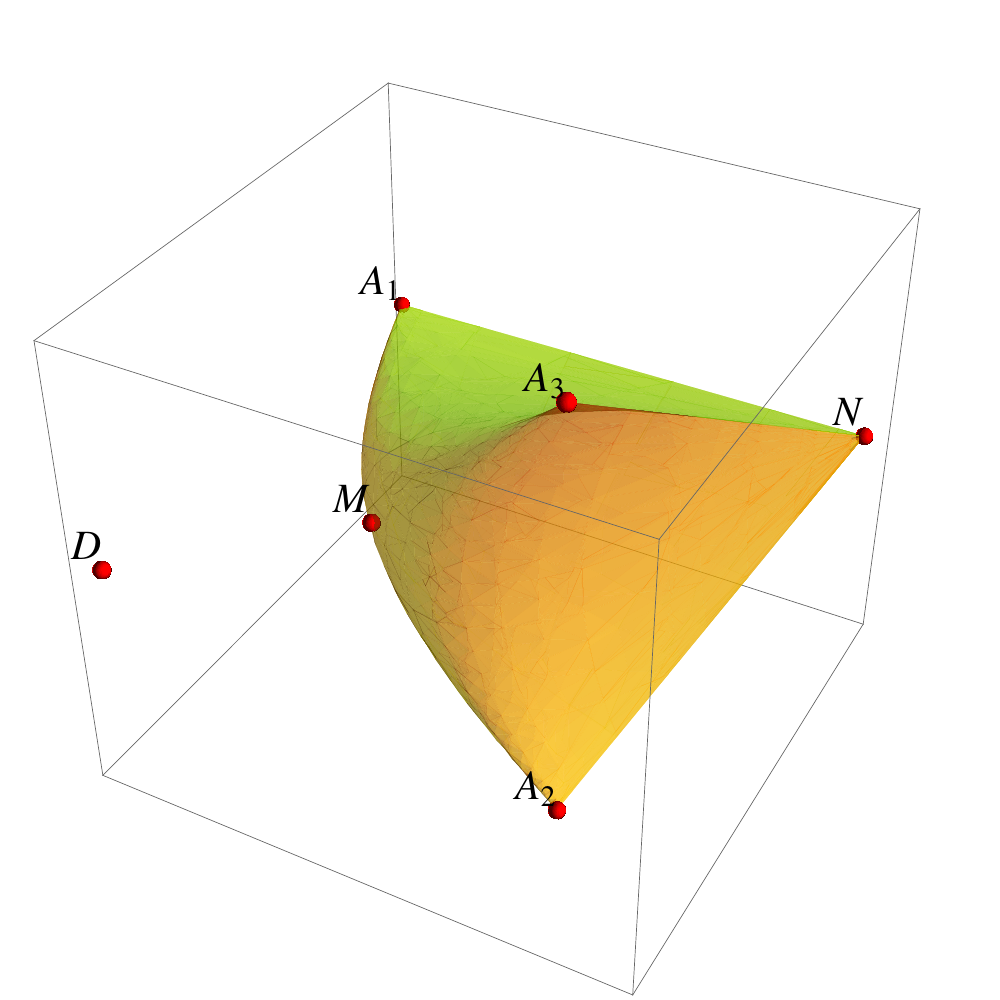} 
\caption{\label{fig:d-body} (Color online)
The position of the Dirichlet $D$ point is shown in relation to the body of allowed conductances in the asymmetric non-chiral case. The fixed points $N$, $A_{i}$, $M$ are also shown. }
\end{figure}

The biggest difference occurs at even stronger attraction. The third line, $(2K+K_{3})/3 = 4$, in our Fig.\ \ref{fig:PortraitK} intersects the regions  IV, V, VI, 3, 4, 3', 6, 2, 1, I, II, these 
correspond again to the sequence of stable FPs :  
$N$, $\bar A$, $\chi^{\pm}$, ($\chi^{\pm}$+$M$), $M$, ($M+A$), $A$. 
This sequence should be compared to Fig.\ 5(f) in \cite{Hou2012}, which provides the list: 
$N$, $\bar A$, $\chi^{\pm}$, $\bar A$,  $D$, $A$. We see that the bosonization analysis, apart from systematically ignoring the existence of the $M$ point, proposes the appearance of the new fixed point $D$. The  ``Dirichlet'' FP $D$ violates the unitarity of the S-matrix, i.e.\ the conservation of the number of particles; the authors of \cite{Oshikawa2006,Hou2012} explain that this property may be attributed to the formation of superconducting pairs close to the Y-junction.  In particular, the $D$ point is proposed as a stable FP even in the fully symmetric situation, $K=K_{3}$.  Our analysis does not indicate the existence of a FP with the properties of the $D$ point, which lies outside the physically accessible region $\Delta$ indicated in Fig.\ \ref{fig:deltoid} at the point $a=b=-1$, $c=0$.  
The relative position of the $D$ point with respect to the body $B'$ of allowed conductance values, available in the fluxless asymmetric case,  is shown in Fig. \ref{fig:d-body}

It should be stressed again, that the scaling exponents at the points $N$, $\chi$, $A$ obtained in \cite{Hou2012} coincide with those, calculated within our approach earlier and in the present paper. Let us look closer at the argument of the authors \cite{Hou2012}  in favor of the $D$ point. They observe that the pair hopping operator at the FP $A$ becomes formally relevant for large attraction, which means that the RG flow generated by such a perturbation should move the system away from the $A$ point.  At the same time, they show that the perturbations around the proposed $D$ points are irrelevant and the $D$ point is thus stable. They do not actually suggest any RG flow, connecting the FPs $A$ and $D$, because these points are proposed by an ansatz,\cite{Hou2012} rather than determined as zeros of the beta-function of a set of RG equations.  

We observe further that the bosonization approach suggests the existence of the stable FP $D$ at strong attraction, whereas the existence of such a point  is not excluded at weaker attraction, when it only becomes  unstable.  Therefore it should be meaningful to discuss values of conductance,  which exceed the maximum allowed value for one-particle processes, for arbitrary interaction, in particular also for weak interaction. One may then hypothetically find examples of contributions from pair hopping, which may increase the conductance over its maximum value, even though the subsequent renormalization (i.e. higher order corrections) reduces this value to usual one. 

From a formal perspective, contributions to the Kubo formula of conductance involving fermion pair processes are certain vertex corrections, first discussed in the context of superconducting fluctuation contributions. These are the Maki-Thompson and Aslamazov-Larkin corrections \cite{LarkinVarlamov}. We showed, however, in our previous work \cite{Aristov2011a} that vertex corrections vanish at zero temperature in the limit $\omega L \to0$.  It is known 
that the criterion $\omega L \ll 1$ is crucial for obtaining the correct value of the conductance of the clean Luttinger liquid with leads \cite{Maslov1995,Ponomarenko1996,Safi1995}.  
This criterion however does not appear in the analysis of  \cite{Oshikawa2006,Hou2012}, where the presence of leads at $x>L$ is modeled at the end of the calculation by introducing ``contact'' resistances, which may be positive or negative depending on the interaction.  
After this modeling the authors  \cite{Oshikawa2006,Hou2012} conclude that the $D$ point exists also in the presence of leads and is characterized by the above values $a=b=-1$, in violation of the unitarity of the single-particle S-matrix. The proposed picture raises a natural question about the role of the length of the interacting region $L$. In our fermionic analysis small $L$ correspond to the absence of any renormalization.  By contrast, the result in \cite{Oshikawa2006,Hou2012} reduces to the statement, that a sufficiently long finite region around the Y-junction should lead at strong attraction between fermions to the peculiar behavior that a fermion coming from one Fermi-liquid lead is scattered to another lead, and takes with it a partner fermion from a third lead. It is apparently suggested  that the finite region close to the junction acts as a pump. If valid, such a picture should have a precursor at moderate strength of interaction. However, we quoted above the absence of such contributions in the direct perturbative calculations up to the third order in the interaction.  We conclude that the most plausible resolution of the contradiction is that the \emph{ad hoc} procedure of modeling the non-interacting outer leads used by \cite{Oshikawa2006,Hou2012}  is not correct. There is no reason why the ``contact resistances'' should be the same for the ideal (clean) TLL-wire  and wires connected by a junction, as assumed in \cite{Oshikawa2006,Hou2012}.

\section{Conclusion}

In this paper we presented results for the effect of fermion-fermion interaction on the transport properties of three quantum wires joined at a Y-junction. We employed the Luttinger liquid (LL) model of spinless fermions to calculate the S-matrix and the conductances of this system for the case of equal interaction constants $g$ in the main wire and different interaction $g_{3}$ in the third wire. Using a purely fermionic representation we are able to directly describe the physical situation of wires of length L adiabatically attached to reservoirs. We argue that in this case (at least at zero temperature) the scattering is purely elastic (excitation of particle-hole pairs is excluded) and therefore described by the single-particle S-matrix (the S-matrix is strongly renormalized by interaction effects). We calculate the renormalization of the S-matrix from a set of three renormalization group equations for three conductances. The RG-equations are derived by calculating the leading scale dependent contributions to all orders in the interaction. These contributions are obtained by analytically solving a set of integral equations (ladder summation). In our earlier work we have presented arguments suggesting that the selected contributions are the dominant ones, meaning that additional contributions do not change the scaling behavior near the fixed points \cite{Aristov2009}. Additional support for the correctness of our method is coming from the fact that our results agree with known results in all cases where these results are well founded. The relative simplicity of our method allows, however, to consider more complex problems not addressed before, e.g. in the present case anisotropic Y-junctions at strong coupling. As demonstrated in our earlier work, the addition of even small non-symmetric perturbances to a highly symmetric system may change the RG-flow pattern in a qualitative way. For example, a second order contribution to the tunneling from a tip into a quantum wire is found to destabilize the fixed point corresponding to the tip separated from the ideally conducting wire \cite{Aristov2010}. Similarly, a small asymmetry in the interaction constants of tip and wire of a chiral Y-junction leads to the appearance of lines of fixed points \cite{Aristov2012a}. The fully symmetric chiral Y-junction has been studied in great detail by \cite{Oshikawa2006}. These authors found the emergence of a pair of chiral FPs, $\chi^{\pm}$, which we confirm in our present analysis, extending the considerations to anisotropic interactions $g \neq g_{3}$. They also conjectured a new and quite unusual fixed point termed $D$, which in their interpretation involved "Andreev scattering", i.e. the tunneling of fermion pairs. Such processes necessarily involve particle-hole excitations, which we exclude from the beginning (these processes have vanishing phase space in the limit of zero excitation energy). The existence of the new FP was argued in \cite{Oshikawa2006} to be necessary to complete the RG flow pattern. We agree that new fixed points are necessary at large attractive interaction. However, we find in the isotropic case a set of additional unstable chiral FPs, $C^{\pm}$, in the domain of attractive interaction, $K>2$, serving a similar purpose. The emergence of $C^{\pm}$ allows to stabilize the FP $M$. In the anisotropic case a further unstable FP, $Q$, is appearing. In a follow up to \cite{Oshikawa2006} the properties of a fully asymmetric Y-junction have been considered recently \cite{Hou2012}. Our detailed comparison with the results obtained by the latter authors reveals agreement in many aspects of this multifaceted problem. At strong attraction, however, our results differ from theirs in two respects: we do not find their fixed point $D$, which violates the unitarity of the single particle S-matrix, but instead find additional fixed points not discovered by them. 
Finally we like to point out that our formulation is rather general, allowing to derive RG-equations for junctions connecting any number of leads in any asymmetric way (see Eq.(\ref{eq:RGgeneral})). The challenge in that case is to find a suitable parametrization of the S-matrix. Work in this direction for the four-lead junction is in progress.

\acknowledgements

We are grateful to I.V. Gornyi, A.W.W. Ludwig, D.G. Polyakov, and A. Rahmani for useful discussions.   
The work of D.A. was supported by the German-Israeli Foundation (GIF), the
Dynasty foundation, and a BMBF grant. The work of D.A. and P.W. was
supported by the DFG-Center for Functional Nanostructures at KIT. One of us (PW) acknowledges the
support by an ICAM Senior Scientist Fellowship during his stay as visiting
professor at the University of Wisconsin, Madison.

\appendix

\section{S-matrix parametrization and re-phasing}
\label{sec:Smatrix}
 
The traceless Gell-Mann matrices, $\lambda _{j}$, with $j=1,\ldots 8$
are the generators of the SU(3) group and are listed elsewhere. \cite{Aristov2011}
Together with the unit matrix $\lambda _{0}$, they form a basis for representing the $S$-matrix of a three-lead junction. We note  the property $\mbox{Tr}[\lambda_{j}\lambda _{k}]=2\delta _{jk}$, with $j,k=0,\ldots 8$.  

Up to an overall phase, an arbitrary matrix $S$, belonging to representations of the SU(3) group, may be defined by its Euler angles parametrization  as follows \cite{Aristov2011} :
\begin{equation}
\begin{aligned}
S &= U e^{i \lambda_{5} \theta}  \bar U  e^{i \lambda_{8}  \alpha_{4} (\sqrt{3}/2)}  , \\
U & = e^{i \lambda_{3}  \alpha_{2}/2}e^{i \lambda_{2} \xi/2}e^{i \lambda_{3} \psi/2}, \\
 \bar U & = 
e^{i \lambda_{3}  \bar\phi/2}e^{i \lambda_{2}  \bar\xi/2}
e^{i \lambda_{3}  \alpha_{3}/2}.
\end{aligned}
\label{Euler1}
\end{equation}
While this expression for $S$ contains eight Euler angles, for our purposes it is convenient to introduce an additional degree of freedom by adding a ninth  angle and defining 
\begin{equation}
S_{1} = e^{i \lambda_{8}  \alpha_{1} (\sqrt{3}/2)} S \,.
\end{equation}
One may now use the redundancy in the latter representation to derive a more convenient representation of $S$. Using symbolic computer calculations, one can verify that $S_{1}$ is invariant under the following change  
\begin{equation}
\begin{aligned}
\alpha_{1} & \to \alpha_{1} - \beta, \quad  
\alpha_{4}   \to \alpha_{4} + \beta, \\ 
\psi  &  \to \psi  + 3 \beta,  \quad
\bar\phi\to \bar\phi - 3\beta . 
\end{aligned}
\end{equation}
for arbitrary $\beta$. This means that we may set, e.g.,  $\bar\phi =0$, by choosing $\beta = \bar\phi/3$. Doing this, we arrive at an equivalent representation of the $S$-matrix in terms of  eight parameters. This representation contains four re-phasing angles $\alpha_{1}, \ldots \alpha_{4}$. These do not appear in the expressions for the conductances. 
  In the main text  we therefore discarded the re-phasing angles, which leads to a simpler parametrization of the $S$-matrix. In sections \ref{sec:PTcond}-\ref{sec:FPs}, we set $\bar\xi = \pi-\xi$ and work with the three angles $\theta$, $\psi$, $\xi$. In sections \ref{sec:stability},  \ref{sec:anisotropy}, Appendix \ref{sec:appGenAsym} we also refer to the most general case with four parameters,  $\theta$, $\psi$, $\xi$, $\bar\xi$.

\section{Correspondence with other authors}
\label{sec:GammaPhi}

Let us first establish the correspondence with the notation of Oshikawa et al. \cite{Oshikawa2006} in their Appendix A. They obtain the $S$-matrix in the form 
\begin{equation}   S = e^{-2ik} \left[ 1-    e^{ik} { \Gamma} M\right]^{-1}
 \left[ 1-   e^{-ik} { \Gamma} M\right]\,, 
 \label{Smatrix-tightbinding}
\end{equation}
where $k$ is the wave-vector, and $ { \Gamma} $ is the amplitude in the tight-binding Hamiltonian,describing the hopping from one wire to another. The matrix $M$ is given by 
$M = e^{i\phi/3} B^{\dagger} + e^{-i\phi/3} B$,   and 
$B =  \left(
\begin{smallmatrix}
0, &1,&0 \\  0, & 0,& 1 \\ 1, &0, & 0
\end{smallmatrix} \right) $. Here $\phi$ measures chirality, i.e.\ is a normalized flux through the Y-junction. 

For small $ \Gamma$ we have (up to an overall phase) $S\simeq 1+2i \sin k\,  \Gamma M$. 
In the tight-binding, particle-hole symmetric case, considered below, we have   $k\simeq k_{F} = \pi/2$  and $S\simeq 1+2i \Gamma M$.  

We can express $M$ in terms of Gell-Mann matrices $\lambda_{i}$ as 
\begin{equation} M =  \cos \frac\phi 3
 (\lambda _{1}+\lambda_{4}+\lambda _{6})
+ \sin \frac\phi 3 (\lambda _{2}-\lambda_{5}+\lambda _{7})\,.
\label {eq:app2}
\end{equation}
It is  natural to parametrize the $S$-matrix in a minimal way by the expression
$ S = \exp\left[ i\theta M  \right]$, 
where $\theta = 2\Gamma $ in the limit  $\Gamma \ll 1$.  It turns out, however, that the expressions for conductances are simplified using a slightly different parametrization. Instead of the pair $(\theta \cos (\phi/3), \theta \sin (\phi/3))$ in Eq.\ (\ref{eq:app2}) 
it is better to use the pair 
 $(\theta_{1}  , \theta_{2} ) $ with 
 \[ \theta \cos (\phi/3) =\theta_{1}/3 , \quad \theta \sin (\phi/3)=  \theta_{2} /\sqrt{3} \,.\] 
 
After this convention, we arrive at the S-matrix in the form 
\begin{eqnarray}
S &=&\exp \left(  \tfrac i{3}\theta_{1} (\lambda _{1}+\lambda_{4}+\lambda _{6})
+\tfrac i{\sqrt{3}}\theta_{2} (\lambda _{2}-\lambda_{5}+\lambda _{7}) \right) 
\nonumber \\
& = &  e^{-i\theta_{1}/3}
( r  + t_{+}  B +  t_{-} B ^{\dagger} ) 
\nonumber \\ &=& 
 e^{-i\theta_{1}/3}\begin{pmatrix} 
r, &t_{+} , & t_{-}
\\ t_{-}, &r, &t_{+} \\ 
t_{+} & t_{-} & r  \end{pmatrix}
 \label{Smatrix1}
\end{eqnarray}%
with 
\begin{equation}
\begin{aligned} r &= \tfrac13 (e^{i\theta_{1}} +2\cos\theta_{2}),   \\t_{\pm} &=
\tfrac13 (e^{i\theta_{1}} - \cos\theta_{2} \pm \sqrt{3} \sin\theta_{2}) \,.
\end{aligned}
\end{equation}
The connection with parametrization (\ref{Smatrix-tightbinding}) is established by taking 
$\theta_{1,2} \ll 1$ : 
\begin{equation}
\begin{aligned}
\theta_{1} & = 3  \Gamma \cos( \phi/3) , \quad 
\theta_{2}  =\sqrt{3} \Gamma \sin( \phi/3)  , 
\label{thetaGamma}
\end{aligned} 
\end{equation}
here $\Gamma$ denotes the hopping amplitude, and $\phi =\pi n $ is standing for the flux, piercing the junction, with 
$n$ the number of flux quanta.  We have in special cases 
\begin{equation}
\begin{aligned}
 |\theta_{1}| &=3  \Gamma,\; \; \theta_{2} = 0, \qquad n=0 \mod 3\\ 
|\theta_{1}| & = |\theta_{2}| =3  \Gamma/2 , \qquad n=\pm 1 \mod 3\\ 
\end{aligned} 
\end{equation}

The components of conductance, $a=b$, $c$,  Eq.\ (\ref{def:abc}), are given by
\begin{equation}
\begin{aligned}
a & = \tfrac32 |r|^{2}-\tfrac12 = 
\tfrac13 (\cos2\theta_{2}+2\cos\theta_{1}\cos\theta_{2}) \\ 
c & = \tfrac{ \sqrt{3}}2 (|t_{+}|^{2}-|t_{-}|^{2}) =
 \tfrac23 \sin\theta_{2}(\cos\theta_{1} - \cos\theta_{2}) 
\end{aligned} \label{def:ac}
\end{equation}
We note that $a$ and $c$ reach their extremal values at $\cos\theta_{1}=\pm 1$, 
and these values describe the boundary  of a domain of the form 
\begin{equation} 
a  = \tfrac13 (\cos2\theta_{2}+2 \cos\theta_{2}) , \quad  
c  = \tfrac23 \sin\theta_{2}(1 - \cos\theta_{2}) ,
  \label{deltoid}
\end{equation}  
This curve is called deltoid, and is depicted  in Fig.\ \ref{fig:deltoid}.
The Abelian bosonization approach may be applied to the problem when the 
condition $a^{2}+c^{2}=1$ is satisfied \cite{Aristov2011}. One observes that there are only three such points in  Fig.\ \ref{fig:deltoid}, the corners of the deltoid.

The simple relation (\ref{thetaGamma}) holds only for small $\Gamma$. For finite 
values of $\Gamma$ we calculate the conductance components directly from 
(\ref{Smatrix-tightbinding}) and obtain
\begin{equation}
\begin{aligned}
a & =1- \frac  {12 \,\Gamma^{2}(1 + \Gamma^{2})}{
(1+ 3\Gamma^{2})^{2}  + 4\Gamma^{6}\cos^{2}\phi}
 \\ 
c & = - \frac  {8\sqrt{3} \, \Gamma^{3} \sin \phi}{
(1+ 3\Gamma^{2})^{2}  +4 \Gamma^{6}\cos^{2}\phi}
\end{aligned} \label{ac:tight}
\end{equation} 
These expressions correspond to Eq. (91) in \cite{Rahmani2012} after a change $\Gamma\to t$. 
The quantities $(a,c)$ lie within a deltoid curve for arbitrary $\Gamma\in (0,\infty)$, $\phi \in (0,2\pi)$, as shown in Fig.\ \ref{fig:deltoid}. The proper deltoid curve is  given by Eqs. (\ref{ac:tight}) at  $\phi =\pi/2$ and $\Gamma\in (0,\infty)$

It was argued in  \cite{Oshikawa2006} that the free-fermion description of the $S$-matrix is not fully applicable in the interacting case. Instead, it was proposed  to use the bosonization approach and to find an analog of small $ \Gamma$ in the bosonic Hamiltonian. When abandoning the fermionic formulation and working directly with currents (densities), one does not find restrictions on the values of $a$, $c$, except for the condition $a^{2}+c^{2}\le 1$, mentioned above.  It was hence implied in  \cite{Oshikawa2006} that such restrictions, imposed by the unitarity of $S$-matrix, are relaxed in the interacting case. 

For completeness, we quote here the lowest-order RG equations in terms of $\theta_{1,2}$
\begin{equation}
\begin{aligned}
\tfrac{d}{d\Lambda}\theta_{1}& = - g \sin \theta_{1} \cos\theta_{2}  ,\\ 
\tfrac{d}{d\Lambda} \theta_{2}& = -\tfrac 13 g \sin \theta_{2} (\cos\theta_{1} + 2 \cos\theta_{2}) \,.
\end{aligned} 
\end{equation}
The resulting RG flows are shown qualitatively in the two first panels in Fig.\ \ref{fig:RGflows}, obtained for moderate interaction strength, $K=0.7, 1.5$.

The point of maximally open Y-junction, or $M$-point, is given by $a=-1/3$, $c=0$  and is obtained in two cases. First, in the absence of the flux, $\phi =0$, and $\Gamma = 1/\sqrt{2}$, and second in the case of maximum flux $\phi =\pm\pi/2$, and infinitely strong hopping, $\Gamma \to \infty$.   It is worth to compare this observation 
with Fig.\ 12 of Ref.\  \cite{Oshikawa2006}, where the case,   $\phi =\pm\pi/2$,  $\Gamma \to \infty$, was associated not with the $M$ point, but rather with a different  FP labeled  $D$. In our analysis we showed that the $M$ point is the only stable FP for $K>3$, as indicated in Fig.\ \ref{fig:RGflows} above. Taken from this perspective, our results for $K>3$ do not contradict the results obtained in  \cite{Oshikawa2006}, since both $\phi =0$,  $\Gamma = 1/\sqrt{2}$ and $\phi =\pm\pi/2$,  $\Gamma \to \infty$ correspond to the same FP, the $M$ point.  The important distinction is, however, that our analysis describes the renormalization of the conductances fully in terms of the one-particle $S$-matrix, where the $D$ point would be in the unphysical regime. By contrast, we find the new FPs $C^{\pm}$, not discussed  in \cite{Oshikawa2006}.

\section{RG equation for conductances}
\label{sec:App1}

In this section we sketch the derivation of the formula (\ref{eq:gladder}). This derivation closely follows our paper \cite{Aristov2012}, dealing with a junction of two non-equal Luttinger liquid wires. The formalism presented in that paper can be naturally generalized to an arbitrary number of wires, connected by a single junction. 

In the lowest order of perturbation theory the renormalization of the S-matrix and the conductance matrix is proportional to the strength of interaction, Eq.\ (\ref{RG1order}). 
In higher orders of interaction the interaction constant, entering the RG equation for the conductance, is an effective quantity, described by a function which is non-local and includes retardation effects.

\begin{figure}[tbp]
\includegraphics*[width=0.6\columnwidth]{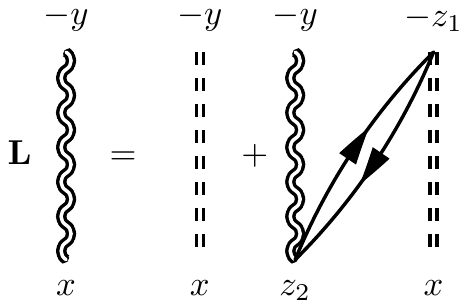}  
\includegraphics*[width=0.7\columnwidth]{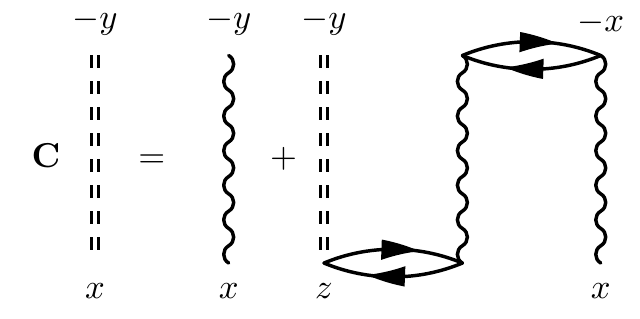}  
\caption{\label{fig:IntegEq} 
Feynman diagrams depicting the integral equations for the renormalized interaction, Eqs.(\ref{Linteq}) and (\ref{Cinteq}). A simple wavy line stands for the initial interaction matrix 
$\mathbf{g}$, the double wavy line defines the quantity $\mathbf{L}$, leading to the effective interaction strength in the RG equations for the conductances.
}
\end{figure}

As shown in \cite{Aristov2009} the leading terms in perturbation theory form a ladder series, which may be summed up analytically to give a function $\mathbf{L}$. The latter obeys the  integral equation: 
 
\begin{equation*}
\begin{aligned} &
\begin{pmatrix} \mathbf{L}(x,y;\omega ) \\   \mathbf{L}_{2}(x,y;\omega )
\end{pmatrix}
=2\pi \mathbf{g}\delta (x-y)
\begin{pmatrix}1 \\ 0 \end{pmatrix}
- 2\pi 
\int_{l}^{L}dz  \\ & \times
\begin{pmatrix} \mathbf{g}   \mathbf{Y} \Pi(x+z,\omega),& 
\mathbf{g}   \Pi(x-z,\omega) \\ 
\mathbf{g}   \Pi(z-x,\omega), & 0   \end{pmatrix}
\begin{pmatrix} \mathbf{L}(z,y;\omega ) \\   \mathbf{L}_{2}(z,y;\omega )
\end{pmatrix} \,,
\end{aligned}
\end{equation*}%
with the fermionic loop $\Pi(x,\omega_{n}) = (2\pi)^{-1} (\delta(x)-|\omega_{n}| 
\theta(x\omega_{n}) e^{-x\omega _{n}})$. 
 
It can be shown, that in the case with different strength of interaction in different semi-wires, it is more convenient to solve this equation in two steps. This is graphically depicted in Fig. \ref{fig:IntegEq}, whereas the details are presented elsewhere. \cite{Aristov2012}

At the first step, the auxiliary quantity $\mathbf{C}$ is introduced as $\mathbf{L}$, taken at $\mathbf{Y}=0$. It satisfies the equation 
\begin{equation}
\begin{aligned}
\mathbf{C}(x,y;\omega )&=2\pi \widetilde{\mathbf{g}}\delta (x-y)
-\tfrac{1}{2}\omega 
\widetilde{\mathbf{g}}\mathbf{g} \int_{l}^{L}dz 
\\ &  \times
\left[e^{-\omega (x+z)}+e^{-\omega |x-z|} \right]\mathbf{C}%
(z,y;\omega ) \,.
\end{aligned}
\label{Cinteq}
\end{equation}
with $\widetilde g_{j} = g_{j} / d_{j}^{2}$ and $d_{j}= \sqrt{1-g_{j}^{2}}$. 
The above condition $\mathbf{Y}=0$ is not intuitively evident, and it describes the absence of correlations between the densities of incoming and outgoing electrons.  In this special case, the effects of scattering are absent and the effects of the interaction amount to the usual dressing of the incoming density by the outgoing one. 
Strictly speaking, the condition $\mathbf{Y}=0$ cannot be realized for any S-matrix, as it violates the charge conservation, and should be viewed as vaguely corresponding to the situation ``in the bulk'', far away from the junction.  

Solving Eq.\ (\ref{Cinteq}),  we then obtain the renormalization of the plasmon velocity and certain prefactors of the interaction strength in the individual wires, which eventually become Luttinger coefficients. 

At the second step, we use the expression for $\mathbf{C}$ thus obtained
 in the equation for $\mathbf{L}$, which now includes the knowledge about the scattering $\mathbf{Y}$ : 
 \begin{equation}
 \begin{aligned}
\mathbf{L}(x,y;\omega ) &=\bigskip \mathbf{C}(x,y;\omega )+\frac{\omega }{2\pi 
}\int_{l}^{L}dz_{1}dz_{2}
\\ & \times
\mathbf{C}(x,z_{1};\omega )\mathbf{Y}e^{-\omega
(z_{1}+z_{2})}\mathbf{L}(z_{2},y;\omega )\,,
\end{aligned}
\label{Linteq}
\end{equation}
The solution of the above integral equation (\ref{Linteq}) is relatively simple. Substituting the result into the expression for the conductances one observes that retardation and non-locality effects, contained in $\mathbf{L}$,  disappear in the limit $\omega L\to0$. We then arrive at the RG equation containing the  expression for the effective interaction in the form 
$\mathbf{g}^{ladder}=2(\mathbf{Q-Y})^{-1}$. This is Eq.\ (\ref{eq:gladder})  quoted in the main text of the paper. 

\section{\label{sec:appGenAsym} Fully asymmetric Y-junction}
 In the most general case the $S$ matrix,  (\ref{Euler1}),  is defined by eight parameters,   but only four parameters,  $\theta$, $\psi$, $\xi$, $\bar\xi$ define the conductance. The parameters of the conductance matrix (\ref{def:abcd}) are given by
\begin{equation}
\begin{aligned}
a&=
\tfrac{\cos^{2} \theta +1}{2}  \cos \xi \cos \bar\xi 
-\cos \theta  \cos \psi  \sin \xi  \sin \bar\xi    \,,\\ 
b&=    \tfrac{1}{2} (3 \cos^{2}  \theta -1)  \,, \\ 
    c &= -\tfrac{\sqrt{3} }{2} \cos \xi  \sin ^2 \theta  
\,, \quad   \bar c = -\tfrac{\sqrt{3} }{2}   \cos \bar\xi \sin ^2\theta   \,.
  \end{aligned}
\end{equation}

The general equation (\ref{eq:RGgeneral}) contains a quantity with four indices,  $F_{jklm} = -\mbox{Tr}\left[\widehat{W}_{jk}^{R}\widehat{W}_{lm}^{R}\right]$. This quantity 
is entirely defined by the parameters of  $S$-matrix, as opposed to the interaction,  entering the definition of $g_{ml}^{ladder,R}$.  It is clear from the properties of Gell-Mann matrices, that $F_{jklm}=0$ if any of its indices equals 3; hence there are only $2^{4}=16$ non-zero components of $F_{jklm}\,$. An analysis similar to the one presented in the main text for the $1-2$ symmetric junction  shows that $F_{jklm}$ can be  expressed through the conductance components, $a$, $b$, $c$, $\bar c$ in a compact way. In view of definition  (\ref{def:abcd}) it is convenient to refer to pairwise combinations of indices as $\{11\}=a$, $\{12\}=c$, $\{21\}=\bar c$, $\{22\}=b$, so that $F_{1121}$ is now denoted as $F_{a\bar c}$ ;  $F_{cb}$ stands for  $F_{1222}$ etc. In this notation we have 
 
\begin{equation}
\begin{aligned}
F_{aa}&=1+b-2 a^2
\,, \quad
F_{bb}= 1+ b -2 b^2 
\,, \\
F_{cc}&=1-b-2 c^2
\,, \quad
F_{\bar c \bar c}=1-b-2 \bar c^2
\,, \\
F_{ab}& = F_{ba}=F_{c\bar c}=F_{\bar cc} =  a (1-b)-c \bar c
\,, \\
F_{ac} &=F_{ca} =\bar c-2 a c
\,, \quad
F_{a\bar c} =F_{\bar ca} =c-2 a \bar c
\,, \\
F_{bc}&=  F_{cb}=-(2 b+1) c
\,, \\
F_{b\bar c}&=F_{\bar cb}=-(2 b+1) \bar c \,, 
\end{aligned}
\end{equation}
These expressions stem from the structure of the SU(3) group, and the reduction to the SU(2) group is obtained by setting $b=1$, $c=\bar c=0$. The latter choice leaves only one non-zero element  $F_{aa}=2-2a^{2}$ in the above formulas and we return to the case of a junction between two Luttinger liquids considered earlier.


\begin{thebibliography}{26}
\expandafter\ifx\csname natexlab\endcsname\relax\def\natexlab#1{#1}\fi
\expandafter\ifx\csname bibnamefont\endcsname\relax
  \def\bibnamefont#1{#1}\fi
\expandafter\ifx\csname bibfnamefont\endcsname\relax
  \def\bibfnamefont#1{#1}\fi
\expandafter\ifx\csname citenamefont\endcsname\relax
  \def\citenamefont#1{#1}\fi
\expandafter\ifx\csname url\endcsname\relax
  \def\url#1{\texttt{#1}}\fi
\expandafter\ifx\csname urlprefix\endcsname\relax\def\urlprefix{URL }\fi
\providecommand{\bibinfo}[2]{#2}
\providecommand{\eprint}[2][]{\url{#2}}

\bibitem[{\citenamefont{Kane and Fisher}(1992)}]{Kane1992}
\bibinfo{author}{\bibfnamefont{C.~L.} \bibnamefont{Kane}} \bibnamefont{and}
  \bibinfo{author}{\bibfnamefont{M.~P.~A.} \bibnamefont{Fisher}},
  \bibinfo{journal}{Phys. Rev. B} \textbf{\bibinfo{volume}{46}},
  \bibinfo{pages}{15233} (\bibinfo{year}{1992}).

\bibitem[{\citenamefont{Furusaki and Nagaosa}(1993)}]{Furusaki1993}
\bibinfo{author}{\bibfnamefont{A.}~\bibnamefont{Furusaki}} \bibnamefont{and}
  \bibinfo{author}{\bibfnamefont{N.}~\bibnamefont{Nagaosa}},
  \bibinfo{journal}{Phys. Rev. B} \textbf{\bibinfo{volume}{47}},
  \bibinfo{pages}{4631} (\bibinfo{year}{1993}).

\bibitem[{\citenamefont{Giamarchi}(2003)}]{GiamarchiBook}
\bibinfo{author}{\bibfnamefont{T.}~\bibnamefont{Giamarchi}},
  \emph{\bibinfo{title}{Quantum Physics in One Dimension}}
  (\bibinfo{publisher}{Clarendon Press}, \bibinfo{address}{Oxford},
  \bibinfo{year}{2003}).

\bibitem[{\citenamefont{Weiss et~al.}(1995)\citenamefont{Weiss, Egger, and
  Sassetti}}]{Weiss1995}
\bibinfo{author}{\bibfnamefont{U.}~\bibnamefont{Weiss}},
  \bibinfo{author}{\bibfnamefont{R.}~\bibnamefont{Egger}}, \bibnamefont{and}
  \bibinfo{author}{\bibfnamefont{M.}~\bibnamefont{Sassetti}},
  \bibinfo{journal}{Phys. Rev. B} \textbf{\bibinfo{volume}{52}},
  \bibinfo{pages}{16707} (\bibinfo{year}{1995}).

\bibitem[{\citenamefont{Fendley et~al.}(1995)\citenamefont{Fendley, Ludwig, and
  Saleur}}]{Fendley1995}
\bibinfo{author}{\bibfnamefont{P.}~\bibnamefont{Fendley}},
  \bibinfo{author}{\bibfnamefont{A.~W.~W.} \bibnamefont{Ludwig}},
  \bibnamefont{and} \bibinfo{author}{\bibfnamefont{H.}~\bibnamefont{Saleur}},
  \bibinfo{journal}{Phys. Rev. B} \textbf{\bibinfo{volume}{52}},
  \bibinfo{pages}{8934} (\bibinfo{year}{1995}).

\bibitem[{\citenamefont{Yue et~al.}(1994)\citenamefont{Yue, Glazman, and
  Matveev}}]{Yue1994}
\bibinfo{author}{\bibfnamefont{D.}~\bibnamefont{Yue}},
  \bibinfo{author}{\bibfnamefont{L.~I.} \bibnamefont{Glazman}},
  \bibnamefont{and} \bibinfo{author}{\bibfnamefont{K.~A.}
  \bibnamefont{Matveev}}, \bibinfo{journal}{Phys. Rev. B}
  \textbf{\bibinfo{volume}{49}}, \bibinfo{pages}{1966} (\bibinfo{year}{1994}).

\bibitem[{\citenamefont{Maslov and Stone}(1995)}]{Maslov1995}
\bibinfo{author}{\bibfnamefont{D.~L.} \bibnamefont{Maslov}} \bibnamefont{and}
  \bibinfo{author}{\bibfnamefont{M.}~\bibnamefont{Stone}},
  \bibinfo{journal}{Phys. Rev. B} \textbf{\bibinfo{volume}{52}},
  \bibinfo{pages}{R5539} (\bibinfo{year}{1995}).

\bibitem[{\citenamefont{Ponomarenko}(1996)}]{Ponomarenko1996}
\bibinfo{author}{\bibfnamefont{V.~V.} \bibnamefont{Ponomarenko}},
  \bibinfo{journal}{Phys. Rev. B} \textbf{\bibinfo{volume}{54}},
  \bibinfo{pages}{10328} (\bibinfo{year}{1996}).

\bibitem[{\citenamefont{Safi and Schulz}(1995)}]{Safi1995}
\bibinfo{author}{\bibfnamefont{I.}~\bibnamefont{Safi}} \bibnamefont{and}
  \bibinfo{author}{\bibfnamefont{H.~J.} \bibnamefont{Schulz}},
  \bibinfo{journal}{Phys. Rev. B} \textbf{\bibinfo{volume}{52}},
  \bibinfo{pages}{R17040} (\bibinfo{year}{1995}).

\bibitem[{\citenamefont{Oshikawa et~al.}(2006)\citenamefont{Oshikawa, Chamon,
  and Affleck}}]{Oshikawa2006}
\bibinfo{author}{\bibfnamefont{M.}~\bibnamefont{Oshikawa}},
  \bibinfo{author}{\bibfnamefont{C.}~\bibnamefont{Chamon}}, \bibnamefont{and}
  \bibinfo{author}{\bibfnamefont{I.}~\bibnamefont{Affleck}},
  \bibinfo{journal}{J. Stat. Mech.} \textbf{\bibinfo{volume}{2006}},
  \bibinfo{pages}{P02008} (\bibinfo{year}{2006}).

\bibitem[{\citenamefont{Polyakov and Gornyi}(2003)}]{Polyakov2003}
\bibinfo{author}{\bibfnamefont{D.~G.} \bibnamefont{Polyakov}} \bibnamefont{and}
  \bibinfo{author}{\bibfnamefont{I.~V.} \bibnamefont{Gornyi}},
  \bibinfo{journal}{Phys. Rev. B} \textbf{\bibinfo{volume}{68}},
  \bibinfo{pages}{035421} (\bibinfo{year}{2003}).

\bibitem[{\citenamefont{Das et~al.}(2004)\citenamefont{Das, Rao, and
  Sen}}]{Das2004}
\bibinfo{author}{\bibfnamefont{S.}~\bibnamefont{Das}},
  \bibinfo{author}{\bibfnamefont{S.}~\bibnamefont{Rao}}, \bibnamefont{and}
  \bibinfo{author}{\bibfnamefont{D.}~\bibnamefont{Sen}},
  \bibinfo{journal}{Phys. Rev. B} \textbf{\bibinfo{volume}{70}},
  \bibinfo{pages}{085318} (\bibinfo{year}{2004}).

\bibitem[{\citenamefont{Lal et~al.}(2002)\citenamefont{Lal, Rao, and
  Sen}}]{Lal2002}
\bibinfo{author}{\bibfnamefont{S.}~\bibnamefont{Lal}},
  \bibinfo{author}{\bibfnamefont{S.}~\bibnamefont{Rao}}, \bibnamefont{and}
  \bibinfo{author}{\bibfnamefont{D.}~\bibnamefont{Sen}},
  \bibinfo{journal}{Phys. Rev. B} \textbf{\bibinfo{volume}{66}},
  \bibinfo{pages}{165327} (\bibinfo{year}{2002}).

\bibitem[{\citenamefont{Devillard et~al.}(2008)\citenamefont{Devillard,
  Gasparian, and Martin}}]{Devillard2008}
\bibinfo{author}{\bibfnamefont{P.}~\bibnamefont{Devillard}},
  \bibinfo{author}{\bibfnamefont{V.}~\bibnamefont{Gasparian}},
  \bibnamefont{and} \bibinfo{author}{\bibfnamefont{T.}~\bibnamefont{Martin}},
  \bibinfo{journal}{Phys. Rev. B} \textbf{\bibinfo{volume}{78}},
  \bibinfo{pages}{085130} (\bibinfo{year}{2008}).

\bibitem[{\citenamefont{Teo and Kane}(2009)}]{Teo2009}
\bibinfo{author}{\bibfnamefont{J.~C.~Y.} \bibnamefont{Teo}} \bibnamefont{and}
  \bibinfo{author}{\bibfnamefont{C.~L.} \bibnamefont{Kane}},
  \bibinfo{journal}{Phys. Rev. B} \textbf{\bibinfo{volume}{79}},
  \bibinfo{pages}{235321} (\bibinfo{year}{2009}).

\bibitem[{\citenamefont{Aristov and W\"{o}lfle}(2009)}]{Aristov2009}
\bibinfo{author}{\bibfnamefont{D.~N.} \bibnamefont{Aristov}} \bibnamefont{and}
  \bibinfo{author}{\bibfnamefont{P.}~\bibnamefont{W\"{o}lfle}},
  \bibinfo{journal}{Phys. Rev. B} \textbf{\bibinfo{volume}{80}},
  \bibinfo{eid}{045109}  (\bibinfo{year}{2009}).

\bibitem[{\citenamefont{Chamon et~al.}(2003)\citenamefont{Chamon, Oshikawa, and
  Affleck}}]{Chamon2003}
\bibinfo{author}{\bibfnamefont{C.}~\bibnamefont{Chamon}},
  \bibinfo{author}{\bibfnamefont{M.}~\bibnamefont{Oshikawa}}, \bibnamefont{and}
  \bibinfo{author}{\bibfnamefont{I.}~\bibnamefont{Affleck}},
  \bibinfo{journal}{Phys. Rev. Lett.} \textbf{\bibinfo{volume}{91}},
  \bibinfo{pages}{206403} (\bibinfo{year}{2003}).

\bibitem[{\citenamefont{Barnab\'e-Th\'eriault
  et~al.}(2005)\citenamefont{Barnab\'e-Th\'eriault, Sedeki, Meden, and
  Sch\"onhammer}}]{Barnabe2005}
\bibinfo{author}{\bibfnamefont{X.}~\bibnamefont{Barnab\'e-Th\'eriault}},
  \bibinfo{author}{\bibfnamefont{A.}~\bibnamefont{Sedeki}},
  \bibinfo{author}{\bibfnamefont{V.}~\bibnamefont{Meden}}, \bibnamefont{and}
  \bibinfo{author}{\bibfnamefont{K.}~\bibnamefont{Sch\"onhammer}},
  \bibinfo{journal}{Phys. Rev. Lett.} \textbf{\bibinfo{volume}{94}},
  \bibinfo{pages}{136405} (\bibinfo{year}{2005}).

\bibitem[{\citenamefont{Aristov and
  W\"olfle}(2012{\natexlab{a}})}]{Aristov2012a}
\bibinfo{author}{\bibfnamefont{D.~N.} \bibnamefont{Aristov}} \bibnamefont{and}
  \bibinfo{author}{\bibfnamefont{P.}~\bibnamefont{W\"olfle}},
  \bibinfo{journal}{Phys. Rev. B} \textbf{\bibinfo{volume}{86}},
  \bibinfo{pages}{035137} (\bibinfo{year}{2012}{\natexlab{a}}).

\bibitem[{\citenamefont{Hou et~al.}(2012)\citenamefont{Hou, Rahmani, Feiguin,
  and Chamon}}]{Hou2012}
\bibinfo{author}{\bibfnamefont{C.-Y.} \bibnamefont{Hou}},
  \bibinfo{author}{\bibfnamefont{A.}~\bibnamefont{Rahmani}},
  \bibinfo{author}{\bibfnamefont{A.~E.} \bibnamefont{Feiguin}},
  \bibnamefont{and} \bibinfo{author}{\bibfnamefont{C.}~\bibnamefont{Chamon}},
  \bibinfo{journal}{Phys. Rev. B} \textbf{\bibinfo{volume}{86}},
  \bibinfo{pages}{075451} (\bibinfo{year}{2012}).

\bibitem[{\citenamefont{Aristov et~al.}(2010)\citenamefont{Aristov, Dmitriev,
  Gornyi, Kachorovskii, Polyakov, and W\"olfle}}]{Aristov2010}
\bibinfo{author}{\bibfnamefont{D.~N.} \bibnamefont{Aristov}},
  \bibinfo{author}{\bibfnamefont{A.~P.} \bibnamefont{Dmitriev}},
  \bibinfo{author}{\bibfnamefont{I.~V.} \bibnamefont{Gornyi}},
  \bibinfo{author}{\bibfnamefont{V.~Y.} \bibnamefont{Kachorovskii}},
  \bibinfo{author}{\bibfnamefont{D.~G.} \bibnamefont{Polyakov}},
  \bibnamefont{and} \bibinfo{author}{\bibfnamefont{P.}~\bibnamefont{W\"olfle}},
  \bibinfo{journal}{Phys. Rev. Lett.} \textbf{\bibinfo{volume}{105}},
  \bibinfo{pages}{266404} (\bibinfo{year}{2010}).

\bibitem[{\citenamefont{Aristov and W\"olfle}(2011)}]{Aristov2011a}
\bibinfo{author}{\bibfnamefont{D.~N.} \bibnamefont{Aristov}} \bibnamefont{and}
  \bibinfo{author}{\bibfnamefont{P.}~\bibnamefont{W\"olfle}},
  \bibinfo{journal}{Phys. Rev. B} \textbf{\bibinfo{volume}{84}},
  \bibinfo{pages}{155426} (\bibinfo{year}{2011}).

\bibitem[{\citenamefont{Aristov}(2011)}]{Aristov2011}
\bibinfo{author}{\bibfnamefont{D.~N.} \bibnamefont{Aristov}},
  \bibinfo{journal}{Phys. Rev. B} \textbf{\bibinfo{volume}{83}},
  \bibinfo{pages}{115446} (\bibinfo{year}{2011}).

\bibitem[{\citenamefont{Larkin and Varlamov}(2005)}]{LarkinVarlamov}
\bibinfo{author}{\bibfnamefont{A.}~\bibnamefont{Larkin}} \bibnamefont{and}
  \bibinfo{author}{\bibfnamefont{A.}~\bibnamefont{Varlamov}},
  \emph{\bibinfo{title}{Theory of fluctuations in superconductors}}
  (\bibinfo{publisher}{Oxford University Press}, \bibinfo{year}{2005}).

\bibitem[{\citenamefont{Rahmani et~al.}(2012)\citenamefont{Rahmani, Hou,
  Feiguin, Oshikawa, Chamon, and Affleck}}]{Rahmani2012}
\bibinfo{author}{\bibfnamefont{A.}~\bibnamefont{Rahmani}},
  \bibinfo{author}{\bibfnamefont{C.-Y.} \bibnamefont{Hou}},
  \bibinfo{author}{\bibfnamefont{A.}~\bibnamefont{Feiguin}},
  \bibinfo{author}{\bibfnamefont{M.}~\bibnamefont{Oshikawa}},
  \bibinfo{author}{\bibfnamefont{C.}~\bibnamefont{Chamon}}, \bibnamefont{and}
  \bibinfo{author}{\bibfnamefont{I.}~\bibnamefont{Affleck}},
  \bibinfo{journal}{Phys. Rev. B} \textbf{\bibinfo{volume}{85}},
  \bibinfo{pages}{045120} (\bibinfo{year}{2012}).

\bibitem[{\citenamefont{Aristov and
  W\"olfle}(2012{\natexlab{b}})}]{Aristov2012}
\bibinfo{author}{\bibfnamefont{D.~N.} \bibnamefont{Aristov}} \bibnamefont{and}
  \bibinfo{author}{\bibfnamefont{P.}~\bibnamefont{W\"olfle}},
  \bibinfo{journal}{Lith. J. Phys.} \textbf{\bibinfo{volume}{52}},
  \bibinfo{pages}{89 } (\bibinfo{year}{2012}{\natexlab{b}})

\end{thebibliography}

\end{document}